\documentclass[aip,jcp,numerical]{revtex4-1}

\usepackage{newtxtext,newtxmath,amsmath} 
\usepackage{natbib}
\usepackage{graphicx}	
\usepackage{color}

\newcommand\apj{Astrophys. J.}

\newcommand\apjl{Astrophys. J. Lett.}
\newcommand\aap{Astron. Astroph.}
\newcommand\jcp{J. Chem. Phys.}
\newcommand\jqsrt{J. Quant. Spec. Rad. Trans.}
\newcommand\pra{Phys. Rev. A}
\newcommand\mnras{Monthly Not. Royal Astron. Soc.}

\begin{document}

\title{Quantum nature of molecular vibrational quenching: Water - molecular hydrogen collisions}
\date{\today}

\author{Laurent Wiesenfeld}
\affiliation{Laboratoire Aim\'e-Cotton, CNRS \& Universit\'e Paris-Saclay, Orsay, France}
\email{laurent.wiesenfeld@universite-paris-saclay.fr}

\begin{abstract}
Rates of conversions
of molecular internal energy to and from kinetic energy by means of molecular collision allows to compute collisional line shapes and transport properties of gases. Knowledge of ro-vibrational quenching rates is necessary to connect spectral observations to physical 
properties of warm astrophysical gasses, including exo-atmospheres.  For a system of paramount 
importance in this context,  the vibrational bending mode quenching of H$_{2}$O by H$_{2}$,  we show here that exchange of vibrational to 
rotational and kinetic energy 
remains  a quantum process, despite the large numbers of quantum levels involved and the large vibrational 
energy transfer.   The excitation of the quantized rotor of the projectile  is by far the 
most effective 
ro-vibrational quenching path of water. To do so, we use a fully quantum first principle computation, potential and 
dynamics, converging it at all stages, in a full coupled channel formalisms. We present here rates for the quenching of the first bending mode of ortho-H$_2$O by ortho H$_2$, up to 500 K, in a fully converged coupled channels formalism.
\end{abstract}

\maketitle

Collisional exchange of energy (internal and kinetic) is a fundamental process leading to the equilibrium between 
internal temperatures of molecules and kinetic temperature of the gas in which they are immersed 
\cite{Hartmann:2008aa}. 
There exists many attempts to compute those rates for transport properties, all 
the way from classical computation (recent examples are  \cite{Hellmann:2020aa,BROWN2011565}) to full quantum 
ab initio computations \cite{Dagdigian:2013aa}. While a large body of investigation deals with collisional rates for 
astrophysics and planetary science, as described in the next paragraph, the role of internal to kinetic energy 
exchanges is also crucial in cold molecules physics ($T\sim\mathrm{K\, \text{to}\, mK}$). In an cold molecular gas, energy  exchanges
 between the molecules observed and with the buffer gas,  possibly causing  heating of the 
cold molecules. \cite{Carr:2009aa,Hu:2021aa}

The recent surge in interest  towards elastic
 and inelastic collisions stems mainly from precise atmospheric/astrochemical needs. \cite{Roueff:2013aa,Karman:2018aa,van-der-Tak:2020aa}
 Most data pertaining to the physical and chemical-physical states of gaseous matter in the Universe come from 
atomic or molecular spectroscopy. For decades, most of the quantitative information was gained from the 
rotational lines of ground state molecules, thanks to the high precision, high specificity of rotational spectra in the cm 
to sub-mm spectral regions \cite{Cernicharo:2019aa,Manigand:2020aa}.

Because of newer 
instruments on the ground and thanks to plane- or space-born telescopes, the opacity of the atmosphere to FIR-IR light 
is progressively 
overcome, and opens up new opportunities: Microwave (THz frequencies) to  Infra Red spectra carry information 
otherwise unattainable, on the warm interstellar matter and on planetary/cometary atmospheres in the Solar System 
or in 
exoplanets. Many molecules are thus nowadays observed in their vibrationally excited states, whether by 
rotational spectroscopy within  those states, or else by observing the FIR-IR lines connecting different ro-vibrational levels 
\cite{Greenwood-A.-J.:2019aa}. Among the molecules observed, water is prominent \cite{van-Dishoeck:2021aa}, being the third most abundant (after H$_2$ and CO) and the first 
polyatomic one. The 
relevance of the ro-vibrational quenching of water is further enhanced by the relatively recent 
discovery of masing transition of ortho-water, in the transition $v_{b}=1; 1_{10}\rightarrow1_{01}$, at 
$\nu=658.00$~GHz \cite{Menten:1995aa,Nesterenok:2015aa,Baudry:2018ac}
[Rotational levels of 
water are labelled $j_{k_{a}k_{c}}$; ortho $^1\mathrm{H_2O}$ (nuclear spin triplet) has $k_{a}+k_{c}$ odd; $v_{b}$ is the water bending 
mode quantum 
number] .

The present approach constitutes, for the author's best knowledge, a first   \textsl{ab initio} study of quantum ro-vibrational quenching 
for a polyatomic molecule, that includes rotational states of the projectile. Integrating these new rotational channels increases 
the magnitude of the vibrational quenching rates (and cross sections) by several orders of magnitude, when compared 
with earlier work, either classical \cite{Faure:2007aa} or quantum \cite{Stoecklin:2019aa,Stoecklin:2021aa} . It has always been 
expected  that quantum effects 
would dominate the landscape for  low   collisional energies,  comparable to the minimum of the van der Waals 
potential $|V_{\text{min}}|$  (here, $V_{\text{min}}\sim-250\,\mathrm{cm^{-1}}$)\cite{Drouin:2012aa,Faure:2013aa,Bergeat:2020aa}, but quantum aspects are supposed to disappear progressively as the collision energy increases In a very different context, 
combustion modelling \cite{BROWN2011565}, the same assumption is made that there exists a critical temperature $T^*= \mathrm k 
T/|V_{\text{min}}|$, above which classical picture is sufficient to describe collisional transport.

 We show here that  the very different energy scales of water rotations, molecular hydrogen rotation, and water 
vibration lead to the resurgence of a strong dependance of the cross sections to the quantum levels of H$_2$ and 
H$_2$O involved. This resurgence does not happen for purely rotational energy redistribution: even classical and/or 
statistical approaches \cite{Loreau:2018aa} are relevant when a large number of coupled levels interact.  It is most
significant here, and, to a lesser extent, to collisions involving water and a heavier partner, like N$_2$, which is 
described by a toy model in this work. 

While classical and semi-classical methods are certainly of relevance in rotational quenching, it must  be recalled 
that the semi-classical quantization of an asymmetric rotor is problematic \cite{Faure:2004aa}, at least for $j$ 
small enough so that the two associated levels $j_{k_a,k_c}$ and $j_{k_a-1,k_c+1}$ are clearly distinct.

Computing  collisional coefficients for purely rotational transitions has been a nearly continuous endeavour since the 
pioneering work of Delgarno 
\cite{Arthurs:1960aa,1974ApJ...191..653G,1977CPL:green,1995JChPh.102.6024P,Flower:2000aa,Dubernet:2013aa,molscat:94,hibridon}. 
The scenario is 
always the same, and  it is the one we 
pursue here: \textsl{(i)} Computing \textit{ab initio}\/ the interaction potential $V_{n}(\bf 
R)$ of the 
polar molecule and the projectile ($\bf R$ denotes collectively all necessary coordinates, $n$ are the points where 
computing is performed) , \textsl{(ii)} fitting the $n$ computed points onto a functional form acceptable for 
dynamics, \textsl{(iii)} running quantum dynamics of the collision pair, and 
\textsl{(iv)} computing cross-section as a function of collisional energy $\sigma(E)$ or rates as a function of the 
kinetic 
temperature of the collider gas $k(T)$.

Thanks to the experience gained in recent precise rotational inelastic scattering computations  and of very 
convincing comparisons of theory with several types of experiments \cite{Ziemkiewicz:2012aa,Drouin:2012aa,Bergeat:2020aa}, it is safe to deal with 
ro-vibrational 
collisional excitation with the same types of methods and basis sets that were used previously,  and conduct a fully  
quantum dynamical approach.

 We employ the full dimensional (9 degrees of freedom) 
water-molecular 
hydrogen potential energy surface \cite{Valiron:2008aa} (hereafter denoted val08), $V(R,\Omega, \delta r_{q})$. This surface includes the 5 
intermolecular degrees of freedom (intermolecular distance $R$ and 4 angles $\Omega$ to orient one molecule with 
respect to the other), as well as 4 normal coordinates describing motion around equilibrium position of the molecules, 
$\delta 
r_{q}$, $q=1,\ldots,4$. Details of this surface are in the original paper. Since quantum dynamics is performed 
with a quantum time-independent computation of the \textsf{S}-matrix, we need to fit  the ab initio 
points on the relevant S. Green type of mixed coordinates \cite{1995JChPh.102.6024P}. Molscat code was used \cite{molscat:94}, duly 
modified 
to include vibrational modes in the potential expression. We did not include the modification of the kinetic energy due 
to rovibrational motion \cite{Faure:2005aa}, but restricted ourselves to potential effects, and expressed the potential in coordinates satisfying Eckart conditions. Since we deal only with the first bending 
mode $v_{b}=1$ and low lying  rotational levels in the excited vibrational mode, we made the assumption:
$\left|v_{b};j_{k_{a},k_{c}}\right> = \left|v_{b}\right>\otimes\left|j_{k_{a},k_{c}}\right>_{v_{b}}$. The rotational 
functions are parameterized by the rotational parameters of either $v_{b}=0$ \cite{Kyro:1981aa} or $v_{b}=1$, but they are not
rovibrational functions. We also averaged the dependance of the potential on the vibrational ground state of H$_{2}$.

Time-independent quantum dynamics is  performed here on two coupled sets of potential energy surfaces with  
full dimensional (5D) potentials,  including  H$_{2}$ rotation (denoted by the quantum number $j_{2}$), in a 
converged coupled channels approach. The 
potentials were obtained by computing 
the averages of the 9-D potential : $C_{v^{'}_{b},v^{}_{b}}(R,\omega) = \left<v'_{b}\left|V(R,\Omega, \delta 
r_{q})\right|v^{}_{b}\right>$ 
functions, with $v^{}_{b}, v'_{b}=0,1$ [see val08]. The coupling $\mathsf W(R)$ matrix 
 is thus written blockwise as (with $R$ the intermolecular distance) \cite{manolopoulos86}:
\begin{equation}
\Psi''(R)=  \left[ \begin{pmatrix}
      \mathsf W_{11}(R) &       \mathsf V_{01}(R) \\
             \mathsf V_{01}(R)&       \mathsf W_{00}(R)
   \end{pmatrix} - \begin{pmatrix} k^{\!2}_{\;2} \\  k^{\!2}_{\;1} \end{pmatrix} \right] \Psi(r) \label{eq:W}
\end{equation}
where the $      \mathsf W_{00}(R)$ and $      \mathsf W_{11}(r)$ matrices are formed by using the potential 
 of each vibrational level  (including the diagonal contributions), and the non-diagonal rotational terms, bracketed with he relevant 
spherical 
harmonic functions \cite{1995JChPh.102.6024P},[val08]. 
The $  \mathsf V_{01}(R)$ matrix is also given by duly bracketing the  $C_{v'_{b},v^{}_{b}}(R,\omega) $ function with 
the same spherical 
harmonic functions.

In order to be able to deal with representative (symmetric) matrices of sizes less than $12,000\times12,000 $ (the 
practical limit of OpenMP computing), we further split the computation by treating each total angular 
momentum value $J$ (recoupling $j$, $j_2$,  $\ell$, the orbital angular momentum) and, if necessary, each inversion symmetry, separately. We used a relatively coarse grid of 
total energies ($E_{\text{tot}}$ from threshold to about 3,000\,$\mathrm{cm^{-1}}$), as we are not interested in the 
detailed 
resonance behaviour occurring typically at $ 1 \,\mathrm{cm^{-1}}< E_{\text{collision}}< 
\left|V_{\text{min}}\right|$.

We present  results for ortho $^1$H$_2$O - ortho $^1$H$_2$ collisions. The present analysis could serve as a basis for the modelling the masing transition of water at 658\,GHz \cite{Baudry:2018aa,Menten:1995aa}; also, ortho H$_2$ is more abundant than para H$_2$ at the higher temperatures that we examine.
With a  large rotational basis set we computed all inelastic cross sections with initial levels $v_{b}=1$; 
$j_{k_{a}k_{c}}=1_{01}$  up to $3_{21}$. Rotational basis for water is $j(v=0)\leq 14$; $j(v=1)\leq 6$; only rotational levels with $E\leq 3500\,\mathrm{cm}^{-1}$ are included. Convergence was reached at higher energies for total angular momentum $J=28$ to 32.

\begin{figure}[htbp]
\includegraphics[width=\textwidth]{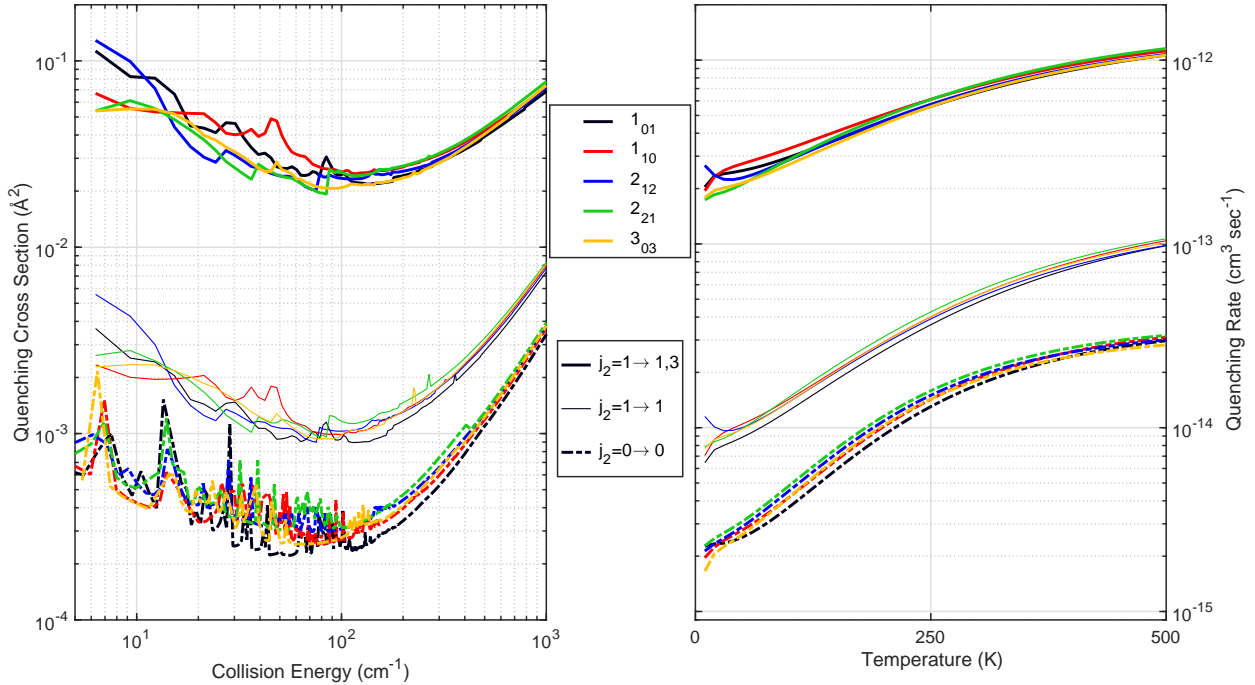}
\caption{\textbf{Left panel: }Quenching cross section (log scale) as a function of collision energy. Individual transitions denoted by colors and line types. 
\textbf{ Right panel}: corresponding rates $k(T)$. Colors of the lines designate  the original $v=1$ rotational level. Line types designate the rotational basis of H$_2$: Lower group: $\sigma^0(E)$; middle group, $\sigma^1(E)$, upper group  $\sigma^{13}(E)$, see text.}
\label{fig:sections}
\end{figure}

Next (Fig.~\ref{fig:sections}), we compute quenching cross sections as a function of scattering energy and quenching 
rates as a function of temperature (inset of Fig.~\ref{fig:sections} ). In both figures, we show the total quenching 
section (or rate)  from a series 
of excited ro-vibrational levels (as described in the caption of Fig. 1) to all $v_b=0$ levels and constraining to 
de-excitation  (Excluding the $v_b=0$ levels that are above the  $v'_b=1$ original levels): 
\begin{equation}
\sigma_{v'=0\leftarrow v=1,j _{k_{a}k_{c}}}(E)=\sum_{j',k'_{a},k'_{c}}\sigma_{v'=0;j' _{k'_{a}k'_{c}}\leftarrow v=1,j 
_{k_{a}k_{c}}}(E) \; ,
\end{equation}
and rates obtained therefrom by averaging over  the collisional 
energy.  Fig.~\ref{fig:sections} shows results for three rotational bases of the H$_{2}$:  (i) para-H$_2$, $j_{2}=0$, to be compared with 
the results of  \cite{Stoecklin:2019aa} (ii) $j_{2}=1$, the usual basis for many computations for collisions between a molecule and   
ortho-H$_{2}$, (iii) a larger basis for ortho-H$_{2}$, $j_{2}=1,3$. Full results with all detailed cross sections and rates will be 
published 
elsewhere. 

Results are striking, as the main channel of quenching appears to go via  a  \emph{simultaneous 
ro-vibration quenching 
of} H$_{2}$O \emph{and a rotational excitation} of H$_{2}$, from $j_{2}=1$ upwards to $j_{2}=3$, an excitation of 
about 600~cm$^{-1}$  (to be compared with a vibrational threshold at 1585~$\mathrm{cm^{-1}}$), 
Fig. \ref{fig:sections}. Note that the $ j_{2}=5$ 
level  has an excitation threshold at 1620~cm$^{-1}$, above the $j_2=1$ levels. We made some preliminary computations of the influence of the full $j_2=1,3,5$ basis. The increase at a collisional energy of about 1500~cm~$^{-1}$ is of the order of 15\%. Full investigation will be presented in future papers.
The relevance of the $\Delta j_{2}=+4$ is marginal at the energies considered here.

Cross sections are very significantly larger for collisions with ortho-H$_{2}$ than with para-H$_{2}$, $j_{2}=0$, an 
expected result , due to the non isotropy of H$_{2}$ (static quadrupole and anisotropic polarizabilty of H$_2$, 
allowing for anisotropic long-distance interaction).  The effect is 
particularly large, but remains compatible  with other collisions with a polar molecule \cite{Yang:2006aa,Daniel:2010aa,Bouhafs:2019aa}. Here, however, 
vibrational quenching cross sections gain another order of magnitude \emph{with including both $j_{2}=1$ and 
$j_{2}=3$ 
ortho rotational levels} of H$_{2}$, remaining with initial conditions at $j_2=1$. An earlier classical work \cite{Faure:2005aa} did not see such a strong effect at all (see their figure 4, with a monotonous decrease of rotational energy of H$_2$ after collision). The  
recent quantum paper \cite{Stoecklin:2019aa}  did not consider these possibilities. Some hints of the importance of the rotational transitions of the projectile are seen in \cite{Yang:2020aa}.

\begin{figure}[htbp]
\begin{center}
\includegraphics[width=0.5\textwidth]{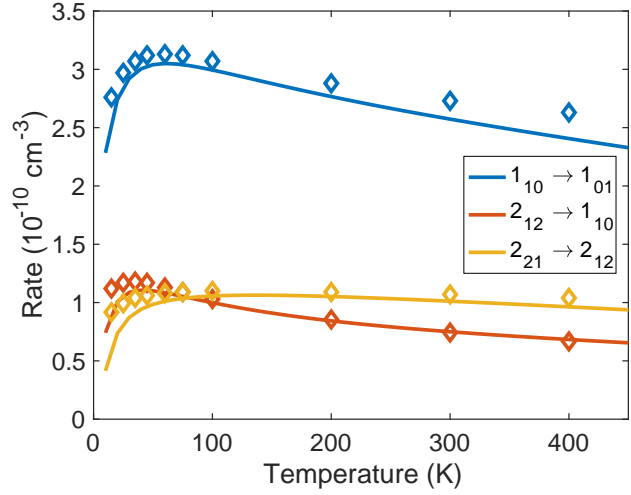}
\caption{ Rates of de-excitation of ortho-H$_2$O levels in $v_b=0$, by ortho-H$_2$.  Diamonds, Daniel et al.\cite{Daniel:2011ab};  line, this work.}
\label{fig:comp}
\end{center}
\end{figure}

Before trying to interpret the surprising results of Fig. \ref{fig:sections}, let us compare those results with previous work. 
We compared
present rate results with earlier ones by Daniel et al., \cite{Daniel:2011ab,Daniel:2011aa}, 
for pure rational quenching inside the $v_{b}=0$ levels, for 
temperatures up to 500~K, see Fig. \ref{fig:comp}.
 Agreement is not perfect, as 
expected, since Daniel et al. had a different definitions of rates and convergence (including the $j_2>0$ channels
 in their initial conditions), unrealistic for the computations here, but the important 
features are present, validating the code and the method.

The $j_2=0$ cross sections found here (lower manifold of  Fig. \ref{fig:sections}) is readily found to be comparable in magnitude to those found by Stoecklin et al., \cite{Stoecklin:2019aa}.  Secondly, in \cite{Faure:2008aa}, a large set of transitions and temperatures was proposed, based on classical and statistical assumptions. It is difficult to compare their results with the ones presented here, as their methodology is more suited for higher $T$, higher initial $j$ and makes no assumption on ortho or para state of H$_2$. To work qualitatively,
 at 300 K, we find a total rate $k(300\,\mathrm K)= 3.5 \,10^{-12} \,\mathrm{cm^3\, sec}$ (summed over all initial levels considered, $1\leq j \leq 3$)  and their rates amounts to $9.7 \,10^{-12}$ approximately, as taken from the LAMDA database, \texttt{https://home.strw.leidenuniv.nl/~moldata/ }(also summed on all initial levels, $j\leq 3$).  Experiment points to $1.3\,10^{-12}$, with no very clear definition of the initial state \cite{Zittel:1991aa}. Clearly, the experiment-theory comparison must be made more precise, probably by examining pressure broadening of IR/Raman spectroscopy.
 
The magnitude of the rates allow us to infer the critical densities. Allowed water IR transitions are the following: $\Delta v_b=\pm 1$; $\Delta j=0,\pm1$, and $\Delta k_a + \Delta k_c=0,\pm2$. IR transitions of the low $j,\, v_b=1$ occur at larger IR frequencies, hence, a relatively large \emph{average} spontaneous emission rate $A=24.6\,\mathrm{ sec^{-1}}$\; \cite{Barber:2006aa}. Critical densities of H$_2$ are thus of the order of $n^*=7\,10^{12}\,\mathrm{cm^{-3}}$, well within the range of atmospheric densities and outflows of aging stars.  Please note that critical densities (as well as rates) are not unambiguously defined here, as many temperatures may coexist. We made use of one kinetic temperature for H$_2$, and summed over all quenching rates. 

To disentangle various possible effects leading to the results of just outlined, we devised four different scenarios  for 
total quenching cross sections $\sigma\equiv\sigma(v'_b=0 \leftarrow v_b=1)$, summed for all  final rotational levels. They differ by the H$_2$ 
rotational 
bases and the way to treat the rotational excitation of H$_2$: \textsl{(i)}  $\sigma^{0}(j_{2}=0)$,  \textsl{(ii)}
 $\sigma^{1} (j_{2}=1)$,  \textsl{(iii)}
$\sigma^{13*} (j_{2}=1,3)$, but forbidding the $\Delta j_{2}=+2$ transitions, and  \textsl{(iv)}  $\sigma^{13} ( j_{2}=1,3)$, allowing 
for 
$\Delta j_{2}=+2$ transitions. They all are shown in Fig.\ref{fig:sections}. Wee see that the $\sigma^{1}$ and $\sigma^{13*}$ cases are very similar,	 with an order of magnitude difference  with $\sigma^{0}$ and $\sigma^{13}$
To understand the dynamics at hand, we took a twofold approach.

\begin{figure}[htbp]
\begin{center}
\includegraphics[width=\textwidth]{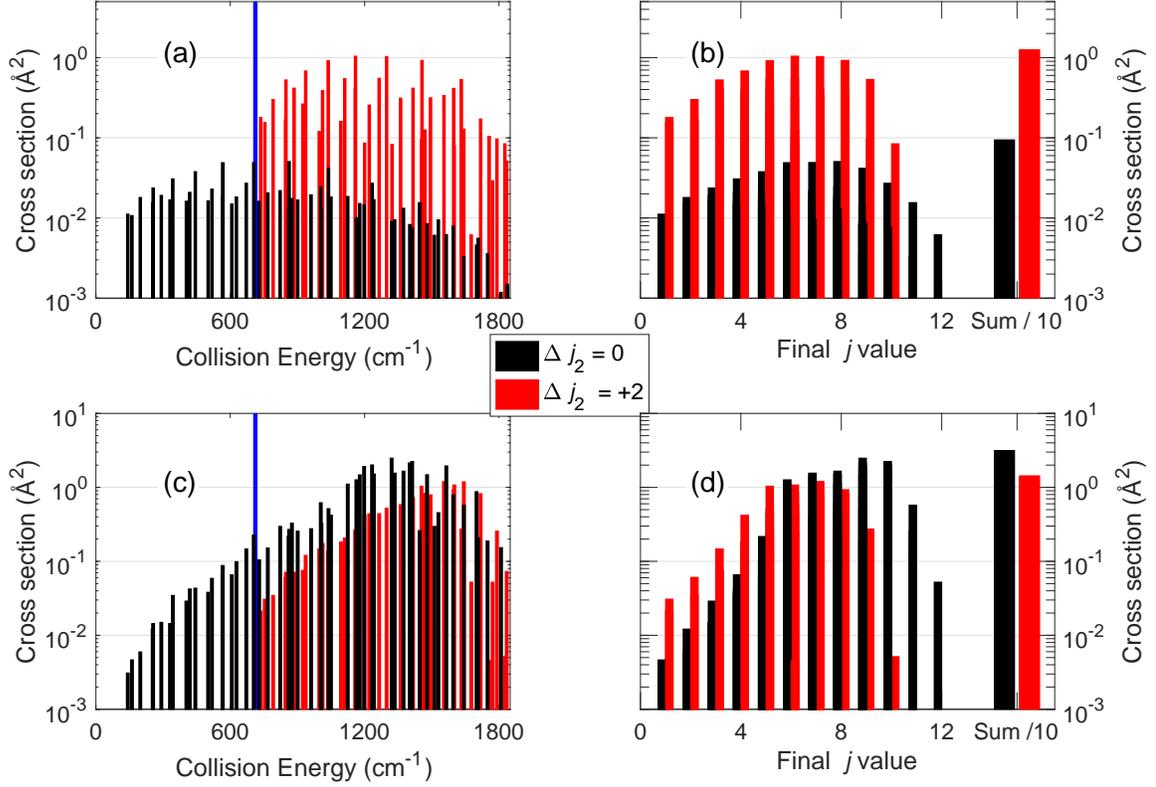}
\caption{ Comparison between rotational quenching, for transition
without excitation of H$_2$ ($\Delta j_2=0$, black bars), and with excitation of H$_2$ ($\Delta j_2=+2$, red bars). Total angular momentum summed, $0\leq J \leq 32$.
Panels (a), (c), cross sections as a function of collision energy. Blue vertical line indicate the $\Delta j_2=+2$ threshold. Panels (b), (d), cross sections as a function of final H$_2$O principal rotation quantum number $j$. $j=11,12$ are above threshold for $\Delta j_2=+2$ excitation.
Panels (a), (b), rovibrational quenching 
 (summed for all initial $ v_b=1,\, j\leq 2$ and all final $v_b=0$, $E_i>E_f$). Panels (c), (d), pure rotational quenching  (summed for all initial $ v_b=0,\, E\geq 1585\,\mathrm{cm}^{-1}$ and all final $v_b=0$, $E_i>E_f$). Last bars in panels (b), (d) sum all contributions.}
\label{fig:histograms}
\end{center}
\end{figure}

\textbf{Firstly,} we compared, for similar quenching energies, the $\Delta j_2=0$ ($\sigma^{13*}$)to the $\Delta j_2=+2$ ($\sigma^{13}$), for ro-vibrational 
quenching or pure rotational quenching (Fig. \ref{fig:histograms}). We present results summed over all partial waves (total angular momentum), either as a function of the total collision energies  
(panels 
(a) and (c)) or as a function of the final rotational state $j$ of (panels(b) and (d)).Initial conditions are chosen as follows:   For panels (a) and (b), initial states are $v=1, j_1=1,2, j_2=0$, that is initial total energy $1737.2<E<  1861.5$ cm$^{-1}$. For panels (c) and (d), we take  initial levels at $v=0$, total $ 1737.2<E<1900$~cm$^{-1}$, and $9\leq j\leq 11$.In this way,  energy gap effects are kept more or less the same, whether with or without vibrational quenching.

Examining the results, for panels (a) and (b) - vibrational quenching-, we see that initial collisional energy has no peculiar feature and that final water angular momentum is weakly peaked at $j\sim6, 7$, for both final $j_2=0,2$. For panels (c) and (d) -no vibrational quenching-, initial collisional energy peaks at about 1300~cm$^{-1}$ and diminishes strongly for smaller values. There is a propensity towards  $\Delta j_2 =+2$ for small final $j$ (large $\Delta j < 0$)
and the opposite for larger final $j$. This goes in favour of a spilt of outgoing angular momentum between the two outgoing products, not seen in vibrational quenching.

The $\Delta 
j_2 = 
+2$ channel dominates the picture for vibrational quenching by at least one order of magnitude when the channels are opened (right of blue 
threshold line, panels (a), (c), $j\leq 10$, panels (b),(d)).  It is not the case for pure rotational quenching. It is all the more remarkable that for the $v=0\rightarrow 0$ 
quenchings, the loss of angular momentum is much higher, with $j_{\text{ini}} \geq 9 $. H$_2$ carrying away angular 
momentum is not the main effect here. Also, energy loss is similar, hence energy gap effects are ruled out in the comparison. Inside the $v_b=0$ vibrational level, we find a weak dominance of the $\Delta j_2 = 0$ transitions, similar to what Daniel et al. found (Figure 1 of reference \citet{Daniel:2010aa})

\par
\begin{figure}[htbp]
\begin{center}
\includegraphics[width=1\textwidth]{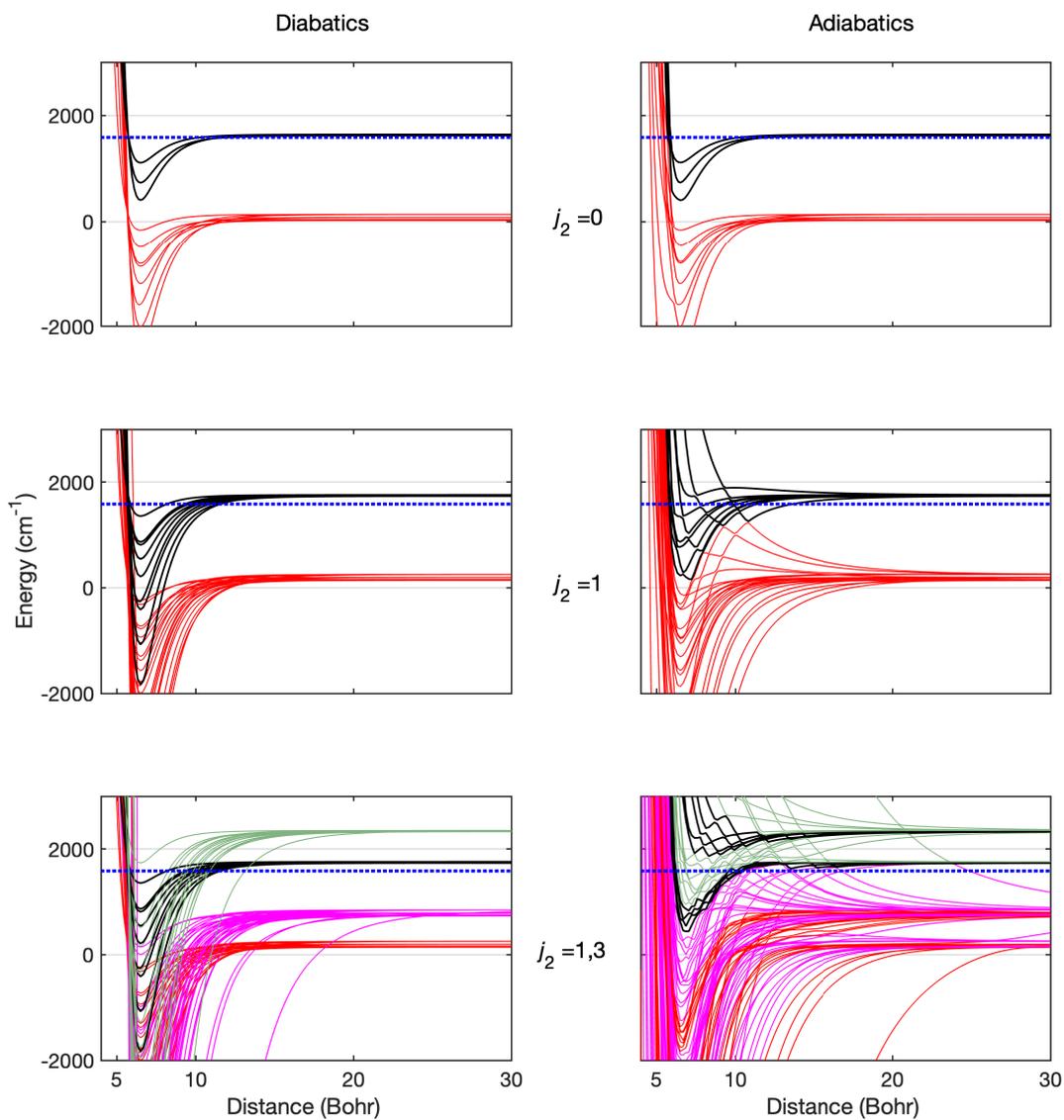}
\caption{ Diabats and adiabats for the $\mathrm {H_2O\cdots H_2}$ collision. Left panels, diabats (diagonal part of the \textsf{W} matrix, Eq.(\ref{eq:W})), as a function of intermolecular distance. Right panels, adiabats (eigenvalues of the \textsf{W} matrix, Eq.(\ref{eq:W})) as a function of intermolecular distance.  Asymptotic quantum numbers : black lines: $v_b=1$, $j_2=0$  or $j_2=1$; red lines $v_b=0$, $j_2=0$  or $j_2=1$; light green  $v_b=1$, $j_2=3$; magenta $v_b=1$, $j_2=1$. Blue dashed line: $v_b=1$, ortho H$_2$O threshold. See discussion in the text.}
\label{fig:dia}
\end{center}
\end{figure}

\textbf{Secondly}, we compute and draw the ro-vibrational channels that are used for the dynamical propagation, as a function of $R$. 
We consider the diagonal part of the  $\mathsf W(R)$ matrix (equation \ref{eq:W}) as a diabatic image and its 
eigenvalues, $\mathrm{eig}\left[ \mathsf W(R)\right]$ as an adiabatic image, Fig. \ref{fig:dia} . Plots are for total $J=2$, 
$j\leq 3$, a good compromise between generality (the $J=6$ or 12 are very similar in appearance; adding $j$ channels do not 
change the picture) and as small a number of channels as possible, for sake of readability. We see straightforwardly that in the case of $j_2=0$, crossings between  incoming black $v_b=1, j_2=0$ 
channels and outgoing red $v_b=0, j_b=0$ channels are limited to the repulsive wall, at very high collision energies. Cross sections are exceedingly small, but increase rapidly once the crossings are open, energy-wise.
Exchanges 
between the various $v_b=0$ channels is possible, seemingly at all energies and distances. The $j_2=1$ case does not 
fundamentally change the picture. There are crossings (diabatic picture) 
or avoided crossings for energies lower than threshold, at $R\sim 8$ Bohr, sufficiently effective to insure some 
transfer of probability amplitude. Note that the higher energy, $R\sim 10$ avoided crossings should be less efficient, 
in a 
Landau-Zener picture, because of the large slope difference between the adiabats. 

Picture changes dramatically 
for the third case, $j=1,3, \, \Delta j_2=0, +2$, two lower panels. If we detail the diabatic  picture (lower left panel), we see (i)
that the $v_b=1$, $j_2=1$ channels (black) are allowed to  cross the  $v_b=1$, $j_2=3$ channels (green) in the $R\sim 10$ region, allowed energy-wise. Then, (ii), the probability amplitude, divided into many channels, crosses a large manifold of $v_b=0, j_2=3$ (magenta) or $v_b=0, j_2=1$ (red) channels, at $R\sim 7-10$, in the diabatic picture, allowing for probability amplitude to flow out on the whole manifold. This picture accounts for the increase of inelastic cross sections and rates,  but without any propensity rule apparent. If we turn to the adiabatic picture (lower right panel), we see that the magenta levels have a series of avoided crossings with the incoming channels at all distances. The larger number of magenta adiabats, their avoided crossings at all distances, including the larger $R$ distances just below threshold may be enough to point to the propensity rules observed in the quantum dynamics.

To conclude the discussion, we found it worthwhile to have a first glimpse of what the situation might be,  for a heavy-heavy collision of utmost atmospheric importance, $\mathrm{H_2O\cdots N_2}$. Even if a very recent rigid bender PES exists, \cite{Wang:2020aa}, fits are not in line of what is needed here, and we look only fore some very qualitative results. We used the same PES as before (same symmetry of the system); it is all the more justified that $V_{\text{min}}$ remains similar (about 250-300 cm$^{-1}$). 
This tN2 (toy-N$_2$) computation of diabats and adiabats is performed exactly similarly, with the two following changes: $B_{\mathrm N_2}=2.01\,\mathrm{cm^{-1}}$ and reduced mass $\mu=10.9565$. While ortho and para $^{14}\mathrm N_2$ have the opposite meaning from $^1\mathrm H_2$ (because of the bosonic character of $^{14}$N, of nuclear spin $I=1$), we did not change the $j_2$ values.

\begin{figure}[htbp]

\includegraphics[width=1\textwidth]{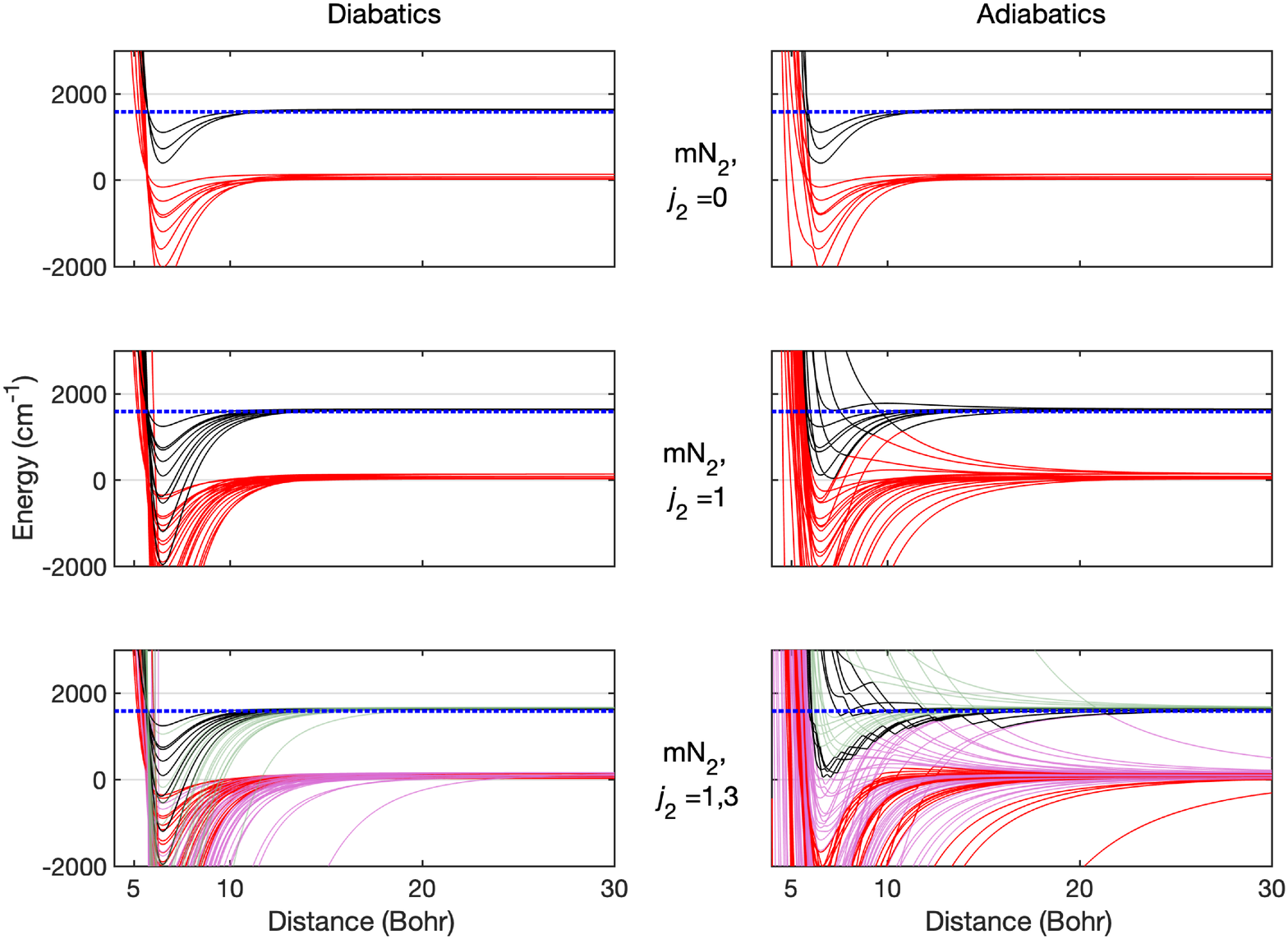}
\begin{center}
\caption{ Diabats and adiabats for the $\mathrm {H_2O\cdots tN_2}$ collision. Left panels, diabats (diagonal part of the \textsf{W} matrix, Eq.(\ref{eq:W})), as a function of intermolecular distance. Right panels, adiabats (eigenvalues of the \textsf{W} matrix, Eq.(\ref{eq:W})) as a function of intermolecular distance.  Asymptotic quantum numbers : black lines: $v_b=1$, $j_2=0$  or $j_2=1$; red lines $v_b=0$, $j_2=0$  or $j_2=1$; light green  $v_b=1$, $j_2=3$; magenta $v_b=1$, $j_2=1$. Blue dashed line: $v_b=1$, ortho H$_2$O threshold. See discussion in the text.}
\label{fig:N2dia}
\end{center}
\end{figure}

 Fig. \ref{fig:N2dia}
show the results, with  the same conventions as in Fig.  \ref{fig:dia}. Without any supposition for the dynamics, we see that the same effects persist, namely, a  large increase of avoided crossings, from $j_2=1$ to $j_2=1,3$.  In this tN2 case we have a large decrease (factor of 25) of the projectile rotational constant (here below the target rotational constants), but the vibrational gap remains. The large fans of adiabats should allow for higher cross sections, because of the various energy scales involved.

 One should not take those tN2 computations for more than indications. In particular, they do not claim for any convergence. However, they should be a warning before deciding whether a cold molecule collision vibrational quenching proceeds classically for all the energy sectors considered. In low $T$ physics, it could be fine structure, vibration, rotation, and each could operate in a very different way.

 Foseca dos Santos et al. \cite{Santos:2011aa} studied some time ago the $\mathrm{H_2\cdots H_2}$ system, considering ro-vibrational de-excitation cross sections, for both ortho-para and ortho-ortho collisions, at mostly low collisions energy (up to 100~K). Since in the present work, molecules are distinguishable, the comparison is more relevant in the ortho-H$_2 \cdots$ para-H$_2$ case, as the ortho-ortho case presents resonances and symmetry effects absent here. We found here that the dominant channel for vibrational quenching is  $j_2=1\rightarrow 3$, irrespective of the collision energy (Fig.~\ref{fig:sections}), thereby reducing the amount of energy to be transferred, independently  of angular momentum transfer. The situation is somewhat different for $\mathrm{H_2 \cdots H_2}$: for the 1001 collision, (notation:$v_1j_1v_2j_2$; 1,2 the two H$_2$ molecules), the dominant channel by far is $1001\rightarrow 0011$, conserving the rotational angular momentum. The analogous $1001 \rightarrow 0201$ is about a factor 5 smaller. An adiabatic analysis similar would sort out the differences, that may be due to the paucity of adiabatic levels (two rods, $B=60\,\mathrm{cm^{-1}}$, far fewer channels). The ortho-ortho case is quite different from the case analyzed here, as resonances and symmetry effects may be dominant effects.

We have  shown that the vibrational quenching behaves in a very different way than the rotational quenching, 
because of the organization of the ro-vibrational quantum levels.  Approximations not allowing rotational (and possibly vibrational) excitation of the projectile may be underestimating the actuel rates by a large amount, here by a factor of 10 at least.  Ignoring the nuclear symmetry of the projectile would also lead to large over-estimation of rates. The results presented here  suggest that ro-vibrational quenching by light atoms or by molecules at their zero angular momentum level is fundamentally different from non-zero angular momentum, because \emph{both} long distance and short distance behaviours of the diabatic or adiabatic channels.

The atmospheric vibrational quenching, driving the IR radiation redistribution towards kinetic energy should thus be carefully conducted in a manner allowing for all  the essential angular momenta to be properly taken into account. Probing  line shapes via IR/Raman spectroscopy is 	all the more relevant and will be the object of further theoretical investigations.The critical density found for ro-vibrational quenching points to high atmospheres of stars or planets, to dense parts of late stage stars, confirming that the 658 GHz maser line finds its origin in the unbalance of collisions and optical rates for the low $j$, $v_b=1$ levels.

As the similarity of rates for the various initial states in our limited range of initial water angular momentum suggests, it is worthwhile to develop specific ways of treating theoretically ro-vibrational quenching, as was  done long ago for rotational quenching, to ease the burden of computation time. In this way, the all-important collision systems for our atmosphere ($\mathrm{H_2O/CO_2/CH_4 \cdots N_2/O_2 }$)will become doable by relevant quantum methods.

\begin{acknowledgements}
 The author thanks J. Hutson and M. Gonzalez-Martinez (Durham U.) for setting up the relevant basis sets for 
vibrational quenching with Molscat conventions, and A. Faure (Grenoble U.) for providing  details of the 
potential surface. Enlightening discussion and support from O. Dulieu and N. Bouloufa are grateful acknowledged. Computations presented in this 
paper were performed with grants  A0060810769/A0080810769 on the ADA and JEAN-ZAY CNRS super-computers, hosted at 
IDRIS.
\end{acknowledgements}

\section*{Data Availability}
The potential energy surface has been published earlier \cite{Valiron:2008aa}. The Fortran subroutines for computing the full surface are available from  the author upon reasonable request.
The quenching data presented here are computed from a large body of raw data, which were generated at the CNRS-IDRIS large facility. Derived data supporting the findings of this study are available from the  author upon reasonable request. 
\bibliography{Vibquench}

\begin{thebibliography}{49}%
\makeatletter
\providecommand \@ifxundefined [1]{%
 \@ifx{#1\undefined}
}%
\providecommand \@ifnum [1]{%
 \ifnum #1\expandafter \@firstoftwo
 \else \expandafter \@secondoftwo
 \fi
}%
\providecommand \@ifx [1]{%
 \ifx #1\expandafter \@firstoftwo
 \else \expandafter \@secondoftwo
 \fi
}%
\providecommand \natexlab [1]{#1}%
\providecommand \enquote  [1]{``#1''}%
\providecommand \bibnamefont  [1]{#1}%
\providecommand \bibfnamefont [1]{#1}%
\providecommand \citenamefont [1]{#1}%
\providecommand \href@noop [0]{\@secondoftwo}%
\providecommand \href [0]{\begingroup \@sanitize@url \@href}%
\providecommand \@href[1]{\@@startlink{#1}\@@href}%
\providecommand \@@href[1]{\endgroup#1\@@endlink}%
\providecommand \@sanitize@url [0]{\catcode `\\12\catcode `\$12\catcode
  `\&12\catcode `\#12\catcode `\^12\catcode `\_12\catcode `\%12\relax}%
\providecommand \@@startlink[1]{}%
\providecommand \@@endlink[0]{}%
\providecommand \url  [0]{\begingroup\@sanitize@url \@url }%
\providecommand \@url [1]{\endgroup\@href {#1}{\urlprefix }}%
\providecommand \urlprefix  [0]{URL }%
\providecommand \Eprint [0]{\href }%
\providecommand \doibase [0]{http://dx.doi.org/}%
\providecommand \selectlanguage [0]{\@gobble}%
\providecommand \bibinfo  [0]{\@secondoftwo}%
\providecommand \bibfield  [0]{\@secondoftwo}%
\providecommand \translation [1]{[#1]}%
\providecommand \BibitemOpen [0]{}%
\providecommand \bibitemStop [0]{}%
\providecommand \bibitemNoStop [0]{.\EOS\space}%
\providecommand \EOS [0]{\spacefactor3000\relax}%
\providecommand \BibitemShut  [1]{\csname bibitem#1\endcsname}%
\let\auto@bib@innerbib\@empty
\bibitem [{\citenamefont {Hartmann}, \citenamefont {Boulet},\ and\
  \citenamefont {Robert}(2008)}]{Hartmann:2008aa}%
  \BibitemOpen
  \bibfield  {author} {\bibinfo {author} {\bibfnamefont {J.}~\bibnamefont
  {Hartmann}}, \bibinfo {author} {\bibfnamefont {C.}~\bibnamefont {Boulet}}, \
  and\ \bibinfo {author} {\bibfnamefont {D.}~\bibnamefont {Robert}},\
  }\href@noop {} {\emph {\bibinfo {title} {Collisional Effects on Molecular
  Spectra}}}\ (\bibinfo  {publisher} {Elsevier},\ \bibinfo {year}
  {2008})\BibitemShut {NoStop}%
\bibitem [{\citenamefont {Hellmann}(2020)}]{Hellmann:2020aa}%
  \BibitemOpen
  \bibfield  {author} {\bibinfo {author} {\bibfnamefont {R.}~\bibnamefont
  {Hellmann}},\ }\href {\doibase 10.1021/acs.jced.0c00605} {\bibfield
  {journal} {\bibinfo  {journal} {Journal of Chemical \& Engineering Data}\
  }\textbf {\bibinfo {volume} {65}},\ \bibinfo {pages} {4712} (\bibinfo {year}
  {2020})},\ \Eprint
  {http://arxiv.org/abs/https://doi.org/10.1021/acs.jced.0c00605}
  {https://doi.org/10.1021/acs.jced.0c00605} \BibitemShut {NoStop}%
\bibitem [{\citenamefont {Brown}, \citenamefont {Bastien},\ and\ \citenamefont
  {Price}(2011)}]{BROWN2011565}%
  \BibitemOpen
  \bibfield  {author} {\bibinfo {author} {\bibfnamefont {N.~J.}\ \bibnamefont
  {Brown}}, \bibinfo {author} {\bibfnamefont {L.~A.}\ \bibnamefont {Bastien}},
  \ and\ \bibinfo {author} {\bibfnamefont {P.~N.}\ \bibnamefont {Price}},\
  }\href {\doibase https://doi.org/10.1016/j.pecs.2010.12.001} {\bibfield
  {journal} {\bibinfo  {journal} {Progress in Energy and Combustion Science}\
  }\textbf {\bibinfo {volume} {37}},\ \bibinfo {pages} {565} (\bibinfo {year}
  {2011})}\BibitemShut {NoStop}%
\bibitem [{\citenamefont {Dagdigian}\ and\ \citenamefont
  {Alexander}(2013)}]{Dagdigian:2013aa}%
  \BibitemOpen
  \bibfield  {author} {\bibinfo {author} {\bibfnamefont {P.~J.}\ \bibnamefont
  {Dagdigian}}\ and\ \bibinfo {author} {\bibfnamefont {M.~H.}\ \bibnamefont
  {Alexander}},\ }\href {\doibase 10.1063/1.4829681} {\bibfield  {journal}
  {\bibinfo  {journal} {The Journal of Chemical Physics}\ }\textbf {\bibinfo
  {volume} {139}},\ \bibinfo {pages} {194309} (\bibinfo {year} {2013})},\
  \Eprint {http://arxiv.org/abs/https://doi.org/10.1063/1.4829681}
  {https://doi.org/10.1063/1.4829681} \BibitemShut {NoStop}%
\bibitem [{\citenamefont {Carr}\ \emph {et~al.}(2009)\citenamefont {Carr},
  \citenamefont {DeMille}, \citenamefont {Krems},\ and\ \citenamefont
  {Ye}}]{Carr:2009aa}%
  \BibitemOpen
  \bibfield  {author} {\bibinfo {author} {\bibfnamefont {L.~D.}\ \bibnamefont
  {Carr}}, \bibinfo {author} {\bibfnamefont {D.}~\bibnamefont {DeMille}},
  \bibinfo {author} {\bibfnamefont {R.~V.}\ \bibnamefont {Krems}}, \ and\
  \bibinfo {author} {\bibfnamefont {J.}~\bibnamefont {Ye}},\ }\href {\doibase
  {10.1088/1367-2630/11/5/055049}} {\bibfield  {journal} {\bibinfo  {journal}
  {{NEW JOURNAL OF PHYSICS}}\ }\textbf {\bibinfo {volume} {{11}}} (\bibinfo
  {year} {{2009}}),\ {10.1088/1367-2630/11/5/055049}}\BibitemShut {NoStop}%
\bibitem [{\citenamefont {Hu}\ \emph {et~al.}(2021)\citenamefont {Hu},
  \citenamefont {Liu}, \citenamefont {Nichols}, \citenamefont {Zhu},
  \citenamefont {Qu{\'e}m{\'e}ner}, \citenamefont {Dulieu},\ and\ \citenamefont
  {Ni}}]{Hu:2021aa}%
  \BibitemOpen
  \bibfield  {author} {\bibinfo {author} {\bibfnamefont {M.-G.}\ \bibnamefont
  {Hu}}, \bibinfo {author} {\bibfnamefont {Y.}~\bibnamefont {Liu}}, \bibinfo
  {author} {\bibfnamefont {M.~A.}\ \bibnamefont {Nichols}}, \bibinfo {author}
  {\bibfnamefont {L.}~\bibnamefont {Zhu}}, \bibinfo {author} {\bibfnamefont
  {G.}~\bibnamefont {Qu{\'e}m{\'e}ner}}, \bibinfo {author} {\bibfnamefont
  {O.}~\bibnamefont {Dulieu}}, \ and\ \bibinfo {author} {\bibfnamefont {K.-K.}\
  \bibnamefont {Ni}},\ }\href {\doibase 10.1038/s41557-020-00610-0} {\bibfield
  {journal} {\bibinfo  {journal} {Nature Chemistry}\ }\textbf {\bibinfo
  {volume} {13}},\ \bibinfo {pages} {435} (\bibinfo {year} {2021})}\BibitemShut
  {NoStop}%
\bibitem [{\citenamefont {{Roueff}}\ and\ \citenamefont
  {{Lique}}(2013)}]{Roueff:2013aa}%
  \BibitemOpen
  \bibfield  {author} {\bibinfo {author} {\bibfnamefont {E.}~\bibnamefont
  {{Roueff}}}\ and\ \bibinfo {author} {\bibfnamefont {F.}~\bibnamefont
  {{Lique}}},\ }\href {\doibase 10.1021/cr400145a} {\bibfield  {journal}
  {\bibinfo  {journal} {Chemical Reviews}\ }\textbf {\bibinfo {volume} {113}},\
  \bibinfo {pages} {8906} (\bibinfo {year} {2013})},\ \Eprint
  {http://arxiv.org/abs/1310.8259} {arXiv:1310.8259 [physics.chem-ph]}
  \BibitemShut {NoStop}%
\bibitem [{\citenamefont {{Karman}}\ \emph {et~al.}(2018)\citenamefont
  {{Karman}}, \citenamefont {{Koenis}}, \citenamefont {{Banerjee}},
  \citenamefont {{Parker}}, \citenamefont {{Gordon}}, \citenamefont {{van der
  Avoird}}, \citenamefont {{van der Zande}},\ and\ \citenamefont
  {{Groenenboom}}}]{Karman:2018aa}%
  \BibitemOpen
  \bibfield  {author} {\bibinfo {author} {\bibfnamefont {T.}~\bibnamefont
  {{Karman}}}, \bibinfo {author} {\bibfnamefont {M.~A.~J.}\ \bibnamefont
  {{Koenis}}}, \bibinfo {author} {\bibfnamefont {A.}~\bibnamefont
  {{Banerjee}}}, \bibinfo {author} {\bibfnamefont {D.~H.}\ \bibnamefont
  {{Parker}}}, \bibinfo {author} {\bibfnamefont {I.~E.}\ \bibnamefont
  {{Gordon}}}, \bibinfo {author} {\bibfnamefont {A.}~\bibnamefont {{van der
  Avoird}}}, \bibinfo {author} {\bibfnamefont {W.~J.}\ \bibnamefont {{van der
  Zande}}}, \ and\ \bibinfo {author} {\bibfnamefont {G.~C.}\ \bibnamefont
  {{Groenenboom}}},\ }\href {\doibase 10.1038/s41557-018-0015-x} {\bibfield
  {journal} {\bibinfo  {journal} {Nature Chemistry}\ }\textbf {\bibinfo
  {volume} {10}},\ \bibinfo {pages} {549} (\bibinfo {year} {2018})}\BibitemShut
  {NoStop}%
\bibitem [{\citenamefont {{van der Tak}}\ \emph {et~al.}(2020)\citenamefont
  {{van der Tak}}, \citenamefont {{Lique}}, \citenamefont {{Faure}},
  \citenamefont {{Black}},\ and\ \citenamefont {{van
  Dishoeck}}}]{van-der-Tak:2020aa}%
  \BibitemOpen
  \bibfield  {author} {\bibinfo {author} {\bibfnamefont {F.~F.~S.}\
  \bibnamefont {{van der Tak}}}, \bibinfo {author} {\bibfnamefont
  {F.}~\bibnamefont {{Lique}}}, \bibinfo {author} {\bibfnamefont
  {A.}~\bibnamefont {{Faure}}}, \bibinfo {author} {\bibfnamefont {J.~H.}\
  \bibnamefont {{Black}}}, \ and\ \bibinfo {author} {\bibfnamefont {E.~F.}\
  \bibnamefont {{van Dishoeck}}},\ }\href {\doibase 10.3390/atoms8020015}
  {\bibfield  {journal} {\bibinfo  {journal} {Atoms}\ }\textbf {\bibinfo
  {volume} {8}},\ \bibinfo {pages} {15} (\bibinfo {year} {2020})},\ \Eprint
  {http://arxiv.org/abs/2004.11230} {arXiv:2004.11230 [astro-ph.GA]}
  \BibitemShut {NoStop}%
\bibitem [{\citenamefont {{Cernicharo}}\ \emph {et~al.}(2019)\citenamefont
  {{Cernicharo}}, \citenamefont {{Gallego}}, \citenamefont
  {{L{\'o}pez-P{\'e}rez}}, \citenamefont {{Tercero}}, \citenamefont
  {{Tanarro}}, \citenamefont {{Beltr{\'a}n}}, \citenamefont {{de Vicente}},
  \citenamefont {{Lauwaet}}, \citenamefont {{Alem{\'a}n}}, \citenamefont
  {{Moreno}}, \citenamefont {{Herrero}}, \citenamefont {{Dom{\'e}nech}},
  \citenamefont {{Ram{\'\i}rez}}, \citenamefont {{Berm{\'u}dez}}, \citenamefont
  {{Pel{\'a}ez}}, \citenamefont {{Patino-Esteban}}, \citenamefont
  {{L{\'o}pez-Fern{\'a}ndez}}, \citenamefont {{Garc{\'\i}a-{\'A}lvaro}},
  \citenamefont {{Garc{\'\i}a-Carre{\~n}o}}, \citenamefont {{Cabezas}},
  \citenamefont {{Malo}}, \citenamefont {{Amils}}, \citenamefont {{Sobrado}},
  \citenamefont {{Diez-Gonz{\'a}lez}}, \citenamefont {{Hernand{\'e}z}},
  \citenamefont {{Tercero}}, \citenamefont {{Santoro}}, \citenamefont
  {{Mart{\'\i}nez}}, \citenamefont {{Castellanos}}, \citenamefont {{Vaquero
  Jim{\'e}nez}}, \citenamefont {{Pardo}}, \citenamefont {{Barbas}},
  \citenamefont {{L{\'o}pez-Fern{\'a}ndez}}, \citenamefont {{Aja}},
  \citenamefont {{Leuther}},\ and\ \citenamefont
  {{Mart{\'\i}n-Gago}}}]{Cernicharo:2019aa}%
  \BibitemOpen
  \bibfield  {author} {\bibinfo {author} {\bibfnamefont {J.}~\bibnamefont
  {{Cernicharo}}}, \bibinfo {author} {\bibfnamefont {J.~D.}\ \bibnamefont
  {{Gallego}}}, \bibinfo {author} {\bibfnamefont {J.~A.}\ \bibnamefont
  {{L{\'o}pez-P{\'e}rez}}}, \bibinfo {author} {\bibfnamefont {F.}~\bibnamefont
  {{Tercero}}}, \bibinfo {author} {\bibfnamefont {I.}~\bibnamefont
  {{Tanarro}}}, \bibinfo {author} {\bibfnamefont {F.}~\bibnamefont
  {{Beltr{\'a}n}}}, \bibinfo {author} {\bibfnamefont {P.}~\bibnamefont {{de
  Vicente}}}, \bibinfo {author} {\bibfnamefont {K.}~\bibnamefont {{Lauwaet}}},
  \bibinfo {author} {\bibfnamefont {B.}~\bibnamefont {{Alem{\'a}n}}}, \bibinfo
  {author} {\bibfnamefont {E.}~\bibnamefont {{Moreno}}}, \bibinfo {author}
  {\bibfnamefont {V.~J.}\ \bibnamefont {{Herrero}}}, \bibinfo {author}
  {\bibfnamefont {J.~L.}\ \bibnamefont {{Dom{\'e}nech}}}, \bibinfo {author}
  {\bibfnamefont {S.~I.}\ \bibnamefont {{Ram{\'\i}rez}}}, \bibinfo {author}
  {\bibfnamefont {C.}~\bibnamefont {{Berm{\'u}dez}}}, \bibinfo {author}
  {\bibfnamefont {R.~J.}\ \bibnamefont {{Pel{\'a}ez}}}, \bibinfo {author}
  {\bibfnamefont {M.}~\bibnamefont {{Patino-Esteban}}}, \bibinfo {author}
  {\bibfnamefont {I.}~\bibnamefont {{L{\'o}pez-Fern{\'a}ndez}}}, \bibinfo
  {author} {\bibfnamefont {S.}~\bibnamefont {{Garc{\'\i}a-{\'A}lvaro}}},
  \bibinfo {author} {\bibfnamefont {P.}~\bibnamefont
  {{Garc{\'\i}a-Carre{\~n}o}}}, \bibinfo {author} {\bibfnamefont
  {C.}~\bibnamefont {{Cabezas}}}, \bibinfo {author} {\bibfnamefont
  {I.}~\bibnamefont {{Malo}}}, \bibinfo {author} {\bibfnamefont
  {R.}~\bibnamefont {{Amils}}}, \bibinfo {author} {\bibfnamefont
  {J.}~\bibnamefont {{Sobrado}}}, \bibinfo {author} {\bibfnamefont
  {C.}~\bibnamefont {{Diez-Gonz{\'a}lez}}}, \bibinfo {author} {\bibfnamefont
  {J.~M.}\ \bibnamefont {{Hernand{\'e}z}}}, \bibinfo {author} {\bibfnamefont
  {B.}~\bibnamefont {{Tercero}}}, \bibinfo {author} {\bibfnamefont
  {G.}~\bibnamefont {{Santoro}}}, \bibinfo {author} {\bibfnamefont
  {L.}~\bibnamefont {{Mart{\'\i}nez}}}, \bibinfo {author} {\bibfnamefont
  {M.}~\bibnamefont {{Castellanos}}}, \bibinfo {author} {\bibfnamefont
  {B.}~\bibnamefont {{Vaquero Jim{\'e}nez}}}, \bibinfo {author} {\bibfnamefont
  {J.~R.}\ \bibnamefont {{Pardo}}}, \bibinfo {author} {\bibfnamefont
  {L.}~\bibnamefont {{Barbas}}}, \bibinfo {author} {\bibfnamefont {J.~A.}\
  \bibnamefont {{L{\'o}pez-Fern{\'a}ndez}}}, \bibinfo {author} {\bibfnamefont
  {B.}~\bibnamefont {{Aja}}}, \bibinfo {author} {\bibfnamefont
  {A.}~\bibnamefont {{Leuther}}}, \ and\ \bibinfo {author} {\bibfnamefont
  {J.~A.}\ \bibnamefont {{Mart{\'\i}n-Gago}}},\ }\href {\doibase
  10.1051/0004-6361/201935197} {\bibfield  {journal} {\bibinfo  {journal}
  {\aap}\ }\textbf {\bibinfo {volume} {626}},\ \bibinfo {eid} {A34} (\bibinfo
  {year} {2019})}\BibitemShut {NoStop}%
\bibitem [{\citenamefont {{Manigand}}\ \emph {et~al.}(2020)\citenamefont
  {{Manigand}}, \citenamefont {{J{\o}rgensen}}, \citenamefont {{Calcutt}},
  \citenamefont {{M{\"u}ller}}, \citenamefont {{Ligterink}}, \citenamefont
  {{Coutens}}, \citenamefont {{Drozdovskaya}}, \citenamefont {{van Dishoeck}},\
  and\ \citenamefont {{Wampfler}}}]{Manigand:2020aa}%
  \BibitemOpen
  \bibfield  {author} {\bibinfo {author} {\bibfnamefont {S.}~\bibnamefont
  {{Manigand}}}, \bibinfo {author} {\bibfnamefont {J.~K.}\ \bibnamefont
  {{J{\o}rgensen}}}, \bibinfo {author} {\bibfnamefont {H.}~\bibnamefont
  {{Calcutt}}}, \bibinfo {author} {\bibfnamefont {H.~S.~P.}\ \bibnamefont
  {{M{\"u}ller}}}, \bibinfo {author} {\bibfnamefont {N.~F.~W.}\ \bibnamefont
  {{Ligterink}}}, \bibinfo {author} {\bibfnamefont {A.}~\bibnamefont
  {{Coutens}}}, \bibinfo {author} {\bibfnamefont {M.~N.}\ \bibnamefont
  {{Drozdovskaya}}}, \bibinfo {author} {\bibfnamefont {E.~F.}\ \bibnamefont
  {{van Dishoeck}}}, \ and\ \bibinfo {author} {\bibfnamefont {S.~F.}\
  \bibnamefont {{Wampfler}}},\ }\href {\doibase 10.1051/0004-6361/201936299}
  {\bibfield  {journal} {\bibinfo  {journal} {\aap}\ }\textbf {\bibinfo
  {volume} {635}},\ \bibinfo {eid} {A48} (\bibinfo {year} {2020})},\ \Eprint
  {http://arxiv.org/abs/2001.06400} {arXiv:2001.06400 [astro-ph.SR]}
  \BibitemShut {NoStop}%
\bibitem [{\citenamefont {{Greenwood, A. J.}}\ \emph
  {et~al.}(2019)\citenamefont {{Greenwood, A. J.}}, \citenamefont {{Kamp, I.}},
  \citenamefont {{Waters, L. B. F. M.}}, \citenamefont {{Woitke, P.}},\ and\
  \citenamefont {{Thi, W.-F.}}}]{Greenwood-A.-J.:2019aa}%
  \BibitemOpen
  \bibfield  {author} {\bibinfo {author} {\bibnamefont {{Greenwood, A. J.}}},
  \bibinfo {author} {\bibnamefont {{Kamp, I.}}}, \bibinfo {author}
  {\bibnamefont {{Waters, L. B. F. M.}}}, \bibinfo {author} {\bibnamefont
  {{Woitke, P.}}}, \ and\ \bibinfo {author} {\bibnamefont {{Thi, W.-F.}}},\
  }\href {\doibase 10.1051/0004-6361/201834175} {\bibfield  {journal} {\bibinfo
   {journal} {A\&A}\ }\textbf {\bibinfo {volume} {631}},\ \bibinfo {pages}
  {A81} (\bibinfo {year} {2019})}\BibitemShut {NoStop}%
\bibitem [{\citenamefont {{van Dishoeck}}\ \emph {et~al.}(2021)\citenamefont
  {{van Dishoeck}}, \citenamefont {{Kristensen}}, \citenamefont {{Mottram}},
  \citenamefont {{Benz}}, \citenamefont {{Bergin}}, \citenamefont {{Caselli}},
  \citenamefont {{Herpin}}, \citenamefont {{Hogerheijde}}, \citenamefont
  {{Johnstone}}, \citenamefont {{Liseau}}, \citenamefont {{Nisini}},
  \citenamefont {{Tafalla}}, \citenamefont {{van der Tak}}, \citenamefont
  {{Wyrowski}}, \citenamefont {{Baudry}}, \citenamefont {{Benedettini}},
  \citenamefont {{Bjerkeli}}, \citenamefont {{Blake}}, \citenamefont
  {{Braine}}, \citenamefont {{Bruderer}}, \citenamefont {{Cabrit}},
  \citenamefont {{Cernicharo}}, \citenamefont {{Choi}}, \citenamefont
  {{Coutens}}, \citenamefont {{de Graauw}}, \citenamefont {{Dominik}},
  \citenamefont {{Fedele}}, \citenamefont {{Fich}}, \citenamefont {{Fuente}},
  \citenamefont {{Furuya}}, \citenamefont {{Goicoechea}}, \citenamefont
  {{Harsono}}, \citenamefont {{Helmich}}, \citenamefont {{Herczeg}},
  \citenamefont {{Jacq}}, \citenamefont {{Karska}}, \citenamefont {{Kaufman}},
  \citenamefont {{Keto}}, \citenamefont {{Lamberts}}, \citenamefont
  {{Larsson}}, \citenamefont {{Leurini}}, \citenamefont {{Lis}}, \citenamefont
  {{Melnick}}, \citenamefont {{Neufeld}}, \citenamefont {{Pagani}},
  \citenamefont {{Persson}}, \citenamefont {{Shipman}}, \citenamefont
  {{Taquet}}, \citenamefont {{van Kempen}}, \citenamefont {{Walsh}},
  \citenamefont {{Wampfler}}, \citenamefont {{Y{\i}ld{\i}z}},\ and\
  \citenamefont {{WISH Team}}}]{van-Dishoeck:2021aa}%
  \BibitemOpen
  \bibfield  {author} {\bibinfo {author} {\bibfnamefont {E.~F.}\ \bibnamefont
  {{van Dishoeck}}}, \bibinfo {author} {\bibfnamefont {L.~E.}\ \bibnamefont
  {{Kristensen}}}, \bibinfo {author} {\bibfnamefont {J.~C.}\ \bibnamefont
  {{Mottram}}}, \bibinfo {author} {\bibfnamefont {A.~O.}\ \bibnamefont
  {{Benz}}}, \bibinfo {author} {\bibfnamefont {E.~A.}\ \bibnamefont
  {{Bergin}}}, \bibinfo {author} {\bibfnamefont {P.}~\bibnamefont {{Caselli}}},
  \bibinfo {author} {\bibfnamefont {F.}~\bibnamefont {{Herpin}}}, \bibinfo
  {author} {\bibfnamefont {M.~R.}\ \bibnamefont {{Hogerheijde}}}, \bibinfo
  {author} {\bibfnamefont {D.}~\bibnamefont {{Johnstone}}}, \bibinfo {author}
  {\bibfnamefont {R.}~\bibnamefont {{Liseau}}}, \bibinfo {author}
  {\bibfnamefont {B.}~\bibnamefont {{Nisini}}}, \bibinfo {author}
  {\bibfnamefont {M.}~\bibnamefont {{Tafalla}}}, \bibinfo {author}
  {\bibfnamefont {F.~F.~S.}\ \bibnamefont {{van der Tak}}}, \bibinfo {author}
  {\bibfnamefont {F.}~\bibnamefont {{Wyrowski}}}, \bibinfo {author}
  {\bibfnamefont {A.}~\bibnamefont {{Baudry}}}, \bibinfo {author}
  {\bibfnamefont {M.}~\bibnamefont {{Benedettini}}}, \bibinfo {author}
  {\bibfnamefont {P.}~\bibnamefont {{Bjerkeli}}}, \bibinfo {author}
  {\bibfnamefont {G.~A.}\ \bibnamefont {{Blake}}}, \bibinfo {author}
  {\bibfnamefont {J.}~\bibnamefont {{Braine}}}, \bibinfo {author}
  {\bibfnamefont {S.}~\bibnamefont {{Bruderer}}}, \bibinfo {author}
  {\bibfnamefont {S.}~\bibnamefont {{Cabrit}}}, \bibinfo {author}
  {\bibfnamefont {J.}~\bibnamefont {{Cernicharo}}}, \bibinfo {author}
  {\bibfnamefont {Y.}~\bibnamefont {{Choi}}}, \bibinfo {author} {\bibfnamefont
  {A.}~\bibnamefont {{Coutens}}}, \bibinfo {author} {\bibfnamefont
  {T.}~\bibnamefont {{de Graauw}}}, \bibinfo {author} {\bibfnamefont
  {C.}~\bibnamefont {{Dominik}}}, \bibinfo {author} {\bibfnamefont
  {D.}~\bibnamefont {{Fedele}}}, \bibinfo {author} {\bibfnamefont
  {M.}~\bibnamefont {{Fich}}}, \bibinfo {author} {\bibfnamefont
  {A.}~\bibnamefont {{Fuente}}}, \bibinfo {author} {\bibfnamefont
  {K.}~\bibnamefont {{Furuya}}}, \bibinfo {author} {\bibfnamefont {J.~R.}\
  \bibnamefont {{Goicoechea}}}, \bibinfo {author} {\bibfnamefont
  {D.}~\bibnamefont {{Harsono}}}, \bibinfo {author} {\bibfnamefont {F.~P.}\
  \bibnamefont {{Helmich}}}, \bibinfo {author} {\bibfnamefont {G.~J.}\
  \bibnamefont {{Herczeg}}}, \bibinfo {author} {\bibfnamefont {T.}~\bibnamefont
  {{Jacq}}}, \bibinfo {author} {\bibfnamefont {A.}~\bibnamefont {{Karska}}},
  \bibinfo {author} {\bibfnamefont {M.}~\bibnamefont {{Kaufman}}}, \bibinfo
  {author} {\bibfnamefont {E.}~\bibnamefont {{Keto}}}, \bibinfo {author}
  {\bibfnamefont {T.}~\bibnamefont {{Lamberts}}}, \bibinfo {author}
  {\bibfnamefont {B.}~\bibnamefont {{Larsson}}}, \bibinfo {author}
  {\bibfnamefont {S.}~\bibnamefont {{Leurini}}}, \bibinfo {author}
  {\bibfnamefont {D.~C.}\ \bibnamefont {{Lis}}}, \bibinfo {author}
  {\bibfnamefont {G.}~\bibnamefont {{Melnick}}}, \bibinfo {author}
  {\bibfnamefont {D.}~\bibnamefont {{Neufeld}}}, \bibinfo {author}
  {\bibfnamefont {L.}~\bibnamefont {{Pagani}}}, \bibinfo {author}
  {\bibfnamefont {M.}~\bibnamefont {{Persson}}}, \bibinfo {author}
  {\bibfnamefont {R.}~\bibnamefont {{Shipman}}}, \bibinfo {author}
  {\bibfnamefont {V.}~\bibnamefont {{Taquet}}}, \bibinfo {author}
  {\bibfnamefont {T.~A.}\ \bibnamefont {{van Kempen}}}, \bibinfo {author}
  {\bibfnamefont {C.}~\bibnamefont {{Walsh}}}, \bibinfo {author} {\bibfnamefont
  {S.~F.}\ \bibnamefont {{Wampfler}}}, \bibinfo {author} {\bibfnamefont
  {U.}~\bibnamefont {{Y{\i}ld{\i}z}}}, \ and\ \bibinfo {author} {\bibnamefont
  {{WISH Team}}},\ }\href {\doibase 10.1051/0004-6361/202039084} {\bibfield
  {journal} {\bibinfo  {journal} {\aap}\ }\textbf {\bibinfo {volume} {648}},\
  \bibinfo {eid} {A24} (\bibinfo {year} {2021})},\ \Eprint
  {http://arxiv.org/abs/2102.02225} {arXiv:2102.02225 [astro-ph.GA]}
  \BibitemShut {NoStop}%
\bibitem [{\citenamefont {{Menten}}\ and\ \citenamefont
  {{Young}}(1995)}]{Menten:1995aa}%
  \BibitemOpen
  \bibfield  {author} {\bibinfo {author} {\bibfnamefont {K.~M.}\ \bibnamefont
  {{Menten}}}\ and\ \bibinfo {author} {\bibfnamefont {K.}~\bibnamefont
  {{Young}}},\ }\href {\doibase 10.1086/316776} {\bibfield  {journal} {\bibinfo
   {journal} {\apjl}\ }\textbf {\bibinfo {volume} {450}},\ \bibinfo {pages}
  {L67} (\bibinfo {year} {1995})}\BibitemShut {NoStop}%
\bibitem [{\citenamefont {{Nesterenok}}(2015)}]{Nesterenok:2015aa}%
  \BibitemOpen
  \bibfield  {author} {\bibinfo {author} {\bibfnamefont {A.~V.}\ \bibnamefont
  {{Nesterenok}}},\ }\href {\doibase 10.1093/mnras/stv485} {\bibfield
  {journal} {\bibinfo  {journal} {\mnras}\ }\textbf {\bibinfo {volume} {449}},\
  \bibinfo {pages} {2875} (\bibinfo {year} {2015})},\ \Eprint
  {http://arxiv.org/abs/1503.07041} {arXiv:1503.07041 [astro-ph.SR]}
  \BibitemShut {NoStop}%
\bibitem [{\citenamefont {{Baudry}}\ \emph
  {et~al.}(2018{\natexlab{a}})\citenamefont {{Baudry}}, \citenamefont
  {{Humphreys}}, \citenamefont {{Herpin}}, \citenamefont {{Torstensson}},
  \citenamefont {{Vlemmings}}, \citenamefont {{Richards}}, \citenamefont
  {{Gray}}, \citenamefont {{De Breuck}},\ and\ \citenamefont
  {{Olberg}}}]{Baudry:2018ac}%
  \BibitemOpen
  \bibfield  {author} {\bibinfo {author} {\bibfnamefont {A.}~\bibnamefont
  {{Baudry}}}, \bibinfo {author} {\bibfnamefont {E.~M.~L.}\ \bibnamefont
  {{Humphreys}}}, \bibinfo {author} {\bibfnamefont {F.}~\bibnamefont
  {{Herpin}}}, \bibinfo {author} {\bibfnamefont {K.}~\bibnamefont
  {{Torstensson}}}, \bibinfo {author} {\bibfnamefont {W.~H.~T.}\ \bibnamefont
  {{Vlemmings}}}, \bibinfo {author} {\bibfnamefont {A.~M.~S.}\ \bibnamefont
  {{Richards}}}, \bibinfo {author} {\bibfnamefont {M.~D.}\ \bibnamefont
  {{Gray}}}, \bibinfo {author} {\bibfnamefont {C.}~\bibnamefont {{De Breuck}}},
  \ and\ \bibinfo {author} {\bibfnamefont {M.}~\bibnamefont {{Olberg}}},\
  }\href {\doibase 10.1051/0004-6361/201731694} {\bibfield  {journal} {\bibinfo
   {journal} {\aap}\ }\textbf {\bibinfo {volume} {609}},\ \bibinfo {eid} {A25}
  (\bibinfo {year} {2018}{\natexlab{a}})},\ \Eprint
  {http://arxiv.org/abs/1711.02350} {arXiv:1711.02350 [astro-ph.SR]}
  \BibitemShut {NoStop}%
\bibitem [{\citenamefont {{Faure}}\ \emph {et~al.}(2007)\citenamefont
  {{Faure}}, \citenamefont {{Crimier}}, \citenamefont {{Ceccarelli}},
  \citenamefont {{Valiron}}, \citenamefont {{Wiesenfeld}},\ and\ \citenamefont
  {{Dubernet}}}]{Faure:2007aa}%
  \BibitemOpen
  \bibfield  {author} {\bibinfo {author} {\bibfnamefont {A.}~\bibnamefont
  {{Faure}}}, \bibinfo {author} {\bibfnamefont {N.}~\bibnamefont {{Crimier}}},
  \bibinfo {author} {\bibfnamefont {C.}~\bibnamefont {{Ceccarelli}}}, \bibinfo
  {author} {\bibfnamefont {P.}~\bibnamefont {{Valiron}}}, \bibinfo {author}
  {\bibfnamefont {L.}~\bibnamefont {{Wiesenfeld}}}, \ and\ \bibinfo {author}
  {\bibfnamefont {M.~L.}\ \bibnamefont {{Dubernet}}},\ }\href {\doibase
  10.1051/0004-6361:20077678} {\bibfield  {journal} {\bibinfo  {journal}
  {\aap}\ }\textbf {\bibinfo {volume} {472}},\ \bibinfo {pages} {1029}
  (\bibinfo {year} {2007})},\ \Eprint {http://arxiv.org/abs/0708.0345}
  {arXiv:0708.0345} \BibitemShut {NoStop}%
\bibitem [{\citenamefont {{Stoecklin}}\ \emph {et~al.}(2019)\citenamefont
  {{Stoecklin}}, \citenamefont {{Denis-Alpizar}}, \citenamefont {{Clergerie}},
  \citenamefont {{Halvick}}, \citenamefont {{Faure}},\ and\ \citenamefont
  {{Scribano}}}]{Stoecklin:2019aa}%
  \BibitemOpen
  \bibfield  {author} {\bibinfo {author} {\bibfnamefont {T.}~\bibnamefont
  {{Stoecklin}}}, \bibinfo {author} {\bibfnamefont {O.}~\bibnamefont
  {{Denis-Alpizar}}}, \bibinfo {author} {\bibfnamefont {A.}~\bibnamefont
  {{Clergerie}}}, \bibinfo {author} {\bibfnamefont {P.}~\bibnamefont
  {{Halvick}}}, \bibinfo {author} {\bibfnamefont {A.}~\bibnamefont {{Faure}}},
  \ and\ \bibinfo {author} {\bibfnamefont {Y.}~\bibnamefont {{Scribano}}},\
  }\href {\doibase 10.1021/acs.jpca.9b04052} {\bibfield  {journal} {\bibinfo
  {journal} {Journal of Physical Chemistry A}\ }\textbf {\bibinfo {volume}
  {123}},\ \bibinfo {pages} {5704} (\bibinfo {year} {2019})}\BibitemShut
  {NoStop}%
\bibitem [{\citenamefont {Stoecklin}\ \emph {et~al.}(2021)\citenamefont
  {Stoecklin}, \citenamefont {Cabrera-Gonz{\'a}lez}, \citenamefont
  {Denis-Alpizar},\ and\ \citenamefont
  {P{\'a}ez-Hern{\'a}ndez}}]{Stoecklin:2021aa}%
  \BibitemOpen
  \bibfield  {author} {\bibinfo {author} {\bibfnamefont {T.}~\bibnamefont
  {Stoecklin}}, \bibinfo {author} {\bibfnamefont {L.~D.}\ \bibnamefont
  {Cabrera-Gonz{\'a}lez}}, \bibinfo {author} {\bibfnamefont {O.}~\bibnamefont
  {Denis-Alpizar}}, \ and\ \bibinfo {author} {\bibfnamefont {D.}~\bibnamefont
  {P{\'a}ez-Hern{\'a}ndez}},\ }\href {\doibase 10.1063/5.0047718} {\bibfield
  {journal} {\bibinfo  {journal} {The Journal of Chemical Physics}\ }\textbf
  {\bibinfo {volume} {154}},\ \bibinfo {pages} {144307} (\bibinfo {year}
  {2021})},\ \Eprint {http://arxiv.org/abs/https://doi.org/10.1063/5.0047718}
  {https://doi.org/10.1063/5.0047718} \BibitemShut {NoStop}%
\bibitem [{\citenamefont {{Drouin}}\ and\ \citenamefont
  {{Wiesenfeld}}(2012)}]{Drouin:2012aa}%
  \BibitemOpen
  \bibfield  {author} {\bibinfo {author} {\bibfnamefont {B.}~\bibnamefont
  {{Drouin}}}\ and\ \bibinfo {author} {\bibfnamefont {L.}~\bibnamefont
  {{Wiesenfeld}}},\ }\href {\doibase 10.1103/PhysRevA.86.022705} {\bibfield
  {journal} {\bibinfo  {journal} {\pra}\ }\textbf {\bibinfo {volume} {86}},\
  \bibinfo {eid} {022705} (\bibinfo {year} {2012})}\BibitemShut {NoStop}%
\bibitem [{\citenamefont {{Faure}}\ \emph {et~al.}(2013)\citenamefont
  {{Faure}}, \citenamefont {{Wiesenfeld}}, \citenamefont {{Drouin}},\ and\
  \citenamefont {{Tennyson}}}]{Faure:2013aa}%
  \BibitemOpen
  \bibfield  {author} {\bibinfo {author} {\bibfnamefont {A.}~\bibnamefont
  {{Faure}}}, \bibinfo {author} {\bibfnamefont {L.}~\bibnamefont
  {{Wiesenfeld}}}, \bibinfo {author} {\bibfnamefont {B.~J.}\ \bibnamefont
  {{Drouin}}}, \ and\ \bibinfo {author} {\bibfnamefont {J.}~\bibnamefont
  {{Tennyson}}},\ }\href {\doibase 10.1016/j.jqsrt.2012.09.015} {\bibfield
  {journal} {\bibinfo  {journal} {\jqsrt}\ }\textbf {\bibinfo {volume} {116}},\
  \bibinfo {pages} {79} (\bibinfo {year} {2013})}\BibitemShut {NoStop}%
\bibitem [{\citenamefont {Bergeat}\ \emph {et~al.}(2020)\citenamefont
  {Bergeat}, \citenamefont {Morales}, \citenamefont {Naulin}, \citenamefont
  {Wiesenfeld},\ and\ \citenamefont {Faure}}]{Bergeat:2020aa}%
  \BibitemOpen
  \bibfield  {author} {\bibinfo {author} {\bibfnamefont {A.}~\bibnamefont
  {Bergeat}}, \bibinfo {author} {\bibfnamefont {S.}~\bibnamefont {Morales}},
  \bibinfo {author} {\bibfnamefont {C.}~\bibnamefont {Naulin}}, \bibinfo
  {author} {\bibfnamefont {L.}~\bibnamefont {Wiesenfeld}}, \ and\ \bibinfo
  {author} {\bibfnamefont {A.}~\bibnamefont {Faure}},\ }\href@noop {}
  {\bibfield  {journal} {\bibinfo  {journal} {Physical Review Letters}\
  }\textbf {\bibinfo {volume} {125}},\ \bibinfo {pages} {143402} (\bibinfo
  {year} {2020})}\BibitemShut {NoStop}%
\bibitem [{\citenamefont {{Loreau}}, \citenamefont {{Lique}},\ and\
  \citenamefont {{Faure}}(2018)}]{Loreau:2018aa}%
  \BibitemOpen
  \bibfield  {author} {\bibinfo {author} {\bibfnamefont {J.}~\bibnamefont
  {{Loreau}}}, \bibinfo {author} {\bibfnamefont {F.}~\bibnamefont {{Lique}}}, \
  and\ \bibinfo {author} {\bibfnamefont {A.}~\bibnamefont {{Faure}}},\ }\href
  {\doibase 10.3847/2041-8213/aaa5fe} {\bibfield  {journal} {\bibinfo
  {journal} {\apjl}\ }\textbf {\bibinfo {volume} {853}},\ \bibinfo {eid} {L5}
  (\bibinfo {year} {2018})},\ \Eprint {http://arxiv.org/abs/1801.02514}
  {arXiv:1801.02514 [astro-ph.IM]} \BibitemShut {NoStop}%
\bibitem [{\citenamefont {{Faure}}\ and\ \citenamefont
  {{Wiesenfeld}}(2004)}]{Faure:2004aa}%
  \BibitemOpen
  \bibfield  {author} {\bibinfo {author} {\bibfnamefont {A.}~\bibnamefont
  {{Faure}}}\ and\ \bibinfo {author} {\bibfnamefont {L.}~\bibnamefont
  {{Wiesenfeld}}},\ }\href {\doibase 10.1063/1.1792613} {\bibfield  {journal}
  {\bibinfo  {journal} {\jcp}\ }\textbf {\bibinfo {volume} {121}},\ \bibinfo
  {pages} {6771} (\bibinfo {year} {2004})}\BibitemShut {NoStop}%
\bibitem [{\citenamefont {{Arthurs}}\ and\ \citenamefont
  {{Dalgarno}}(1960)}]{Arthurs:1960aa}%
  \BibitemOpen
  \bibfield  {author} {\bibinfo {author} {\bibfnamefont {A.~M.}\ \bibnamefont
  {{Arthurs}}}\ and\ \bibinfo {author} {\bibfnamefont {A.}~\bibnamefont
  {{Dalgarno}}},\ }\href {\doibase 10.1098/rspa.1960.0125} {\bibfield
  {journal} {\bibinfo  {journal} {Proceedings of the Royal Society of London
  Series A}\ }\textbf {\bibinfo {volume} {256}},\ \bibinfo {pages} {540}
  (\bibinfo {year} {1960})}\BibitemShut {NoStop}%
\bibitem [{\citenamefont {{Green}}\ and\ \citenamefont
  {{Thaddeus}}(1974)}]{1974ApJ...191..653G}%
  \BibitemOpen
  \bibfield  {author} {\bibinfo {author} {\bibfnamefont {S.}~\bibnamefont
  {{Green}}}\ and\ \bibinfo {author} {\bibfnamefont {P.}~\bibnamefont
  {{Thaddeus}}},\ }\href {\doibase 10.1086/153006} {\bibfield  {journal}
  {\bibinfo  {journal} {\apj}\ }\textbf {\bibinfo {volume} {191}},\ \bibinfo
  {pages} {653} (\bibinfo {year} {1974})}\BibitemShut {NoStop}%
\bibitem [{\citenamefont {{Green}}(1977)}]{1977CPL:green}%
  \BibitemOpen
  \bibfield  {author} {\bibinfo {author} {\bibfnamefont {S.}~\bibnamefont
  {{Green}}},\ }\href@noop {} {\bibfield  {journal} {\bibinfo  {journal}
  {Chemical Physics Letters}\ }\textbf {\bibinfo {volume} {47}},\ \bibinfo
  {pages} {119} (\bibinfo {year} {1977})}\BibitemShut {NoStop}%
\bibitem [{\citenamefont {{Phillips}}, \citenamefont {{Maluendes}},\ and\
  \citenamefont {{Green}}(1995)}]{1995JChPh.102.6024P}%
  \BibitemOpen
  \bibfield  {author} {\bibinfo {author} {\bibfnamefont {T.~R.}\ \bibnamefont
  {{Phillips}}}, \bibinfo {author} {\bibfnamefont {S.}~\bibnamefont
  {{Maluendes}}}, \ and\ \bibinfo {author} {\bibfnamefont {S.}~\bibnamefont
  {{Green}}},\ }\href {\doibase 10.1063/1.469337} {\bibfield  {journal}
  {\bibinfo  {journal} {\jcp}\ }\textbf {\bibinfo {volume} {102}},\ \bibinfo
  {pages} {6024} (\bibinfo {year} {1995})}\BibitemShut {NoStop}%
\bibitem [{\citenamefont {{Flower}}, \citenamefont {{Bourhis}},\ and\
  \citenamefont {{Launay}}(2000)}]{Flower:2000aa}%
  \BibitemOpen
  \bibfield  {author} {\bibinfo {author} {\bibfnamefont {D.~R.}\ \bibnamefont
  {{Flower}}}, \bibinfo {author} {\bibfnamefont {G.}~\bibnamefont {{Bourhis}}},
  \ and\ \bibinfo {author} {\bibfnamefont {J.}~\bibnamefont {{Launay}}},\
  }\href {\doibase 10.1016/S0010-4655(00)00118-1} {\bibfield  {journal}
  {\bibinfo  {journal} {Computer Physics Communications}\ }\textbf {\bibinfo
  {volume} {131}},\ \bibinfo {pages} {187} (\bibinfo {year}
  {2000})}\BibitemShut {NoStop}%
\bibitem [{\citenamefont {{Dubernet}}\ \emph {et~al.}(2013)\citenamefont
  {{Dubernet}}, \citenamefont {{Alexander}}, \citenamefont {{Ba}},
  \citenamefont {{Balakrishnan}}, \citenamefont {{Balan{\c c}a}}, \citenamefont
  {{Ceccarelli}}, \citenamefont {{Cernicharo}}, \citenamefont {{Daniel}},
  \citenamefont {{Dayou}}, \citenamefont {{Doronin}}, \citenamefont
  {{Dumouchel}}, \citenamefont {{Faure}}, \citenamefont {{Feautrier}},
  \citenamefont {{Flower}}, \citenamefont {{Grosjean}}, \citenamefont
  {{Halvick}}, \citenamefont {{K{\l}os}}, \citenamefont {{Lique}},
  \citenamefont {{McBane}}, \citenamefont {{Marinakis}}, \citenamefont
  {{Moreau}}, \citenamefont {{Moszynski}}, \citenamefont {{Neufeld}},
  \citenamefont {{Roueff}}, \citenamefont {{Schilke}}, \citenamefont
  {{Spielfiedel}}, \citenamefont {{Stancil}}, \citenamefont {{Stoecklin}},
  \citenamefont {{Tennyson}}, \citenamefont {{Yang}}, \citenamefont
  {{Vasserot}},\ and\ \citenamefont {{Wiesenfeld}}}]{Dubernet:2013aa}%
  \BibitemOpen
  \bibfield  {author} {\bibinfo {author} {\bibfnamefont {M.-L.}\ \bibnamefont
  {{Dubernet}}}, \bibinfo {author} {\bibfnamefont {M.~H.}\ \bibnamefont
  {{Alexander}}}, \bibinfo {author} {\bibfnamefont {Y.~A.}\ \bibnamefont
  {{Ba}}}, \bibinfo {author} {\bibfnamefont {N.}~\bibnamefont
  {{Balakrishnan}}}, \bibinfo {author} {\bibfnamefont {C.}~\bibnamefont
  {{Balan{\c c}a}}}, \bibinfo {author} {\bibfnamefont {C.}~\bibnamefont
  {{Ceccarelli}}}, \bibinfo {author} {\bibfnamefont {J.}~\bibnamefont
  {{Cernicharo}}}, \bibinfo {author} {\bibfnamefont {F.}~\bibnamefont
  {{Daniel}}}, \bibinfo {author} {\bibfnamefont {F.}~\bibnamefont {{Dayou}}},
  \bibinfo {author} {\bibfnamefont {M.}~\bibnamefont {{Doronin}}}, \bibinfo
  {author} {\bibfnamefont {F.}~\bibnamefont {{Dumouchel}}}, \bibinfo {author}
  {\bibfnamefont {A.}~\bibnamefont {{Faure}}}, \bibinfo {author} {\bibfnamefont
  {N.}~\bibnamefont {{Feautrier}}}, \bibinfo {author} {\bibfnamefont {D.~R.}\
  \bibnamefont {{Flower}}}, \bibinfo {author} {\bibfnamefont {A.}~\bibnamefont
  {{Grosjean}}}, \bibinfo {author} {\bibfnamefont {P.}~\bibnamefont
  {{Halvick}}}, \bibinfo {author} {\bibfnamefont {J.}~\bibnamefont
  {{K{\l}os}}}, \bibinfo {author} {\bibfnamefont {F.}~\bibnamefont {{Lique}}},
  \bibinfo {author} {\bibfnamefont {G.~C.}\ \bibnamefont {{McBane}}}, \bibinfo
  {author} {\bibfnamefont {S.}~\bibnamefont {{Marinakis}}}, \bibinfo {author}
  {\bibfnamefont {N.}~\bibnamefont {{Moreau}}}, \bibinfo {author}
  {\bibfnamefont {R.}~\bibnamefont {{Moszynski}}}, \bibinfo {author}
  {\bibfnamefont {D.~A.}\ \bibnamefont {{Neufeld}}}, \bibinfo {author}
  {\bibfnamefont {E.}~\bibnamefont {{Roueff}}}, \bibinfo {author}
  {\bibfnamefont {P.}~\bibnamefont {{Schilke}}}, \bibinfo {author}
  {\bibfnamefont {A.}~\bibnamefont {{Spielfiedel}}}, \bibinfo {author}
  {\bibfnamefont {P.~C.}\ \bibnamefont {{Stancil}}}, \bibinfo {author}
  {\bibfnamefont {T.}~\bibnamefont {{Stoecklin}}}, \bibinfo {author}
  {\bibfnamefont {J.}~\bibnamefont {{Tennyson}}}, \bibinfo {author}
  {\bibfnamefont {B.}~\bibnamefont {{Yang}}}, \bibinfo {author} {\bibfnamefont
  {A.-M.}\ \bibnamefont {{Vasserot}}}, \ and\ \bibinfo {author} {\bibfnamefont
  {L.}~\bibnamefont {{Wiesenfeld}}},\ }\href {\doibase
  10.1051/0004-6361/201220630} {\bibfield  {journal} {\bibinfo  {journal}
  {\aap}\ }\textbf {\bibinfo {volume} {553}},\ \bibinfo {eid} {A50} (\bibinfo
  {year} {2013})}\BibitemShut {NoStop}%
\bibitem [{\citenamefont {Hutson}\ and\ \citenamefont
  {Green}(1994)}]{molscat:94}%
  \BibitemOpen
  \bibfield  {author} {\bibinfo {author} {\bibfnamefont {J.~M.}\ \bibnamefont
  {Hutson}}\ and\ \bibinfo {author} {\bibfnamefont {S.}~\bibnamefont {Green}},\
  }\href@noop {} {\enquote {\bibinfo {title} {{\sc molscat} computer code,
  version 14 (1994), distributed by collaborative computational project no. 6
  of the engineering and physical sciences research council (uk)},}\ }
  (\bibinfo {year} {1994})\BibitemShut {NoStop}%
\bibitem [{\citenamefont {{M. H. Alexander and D. E. Manolopoulos and H.-J.
  Werner and B. Follmeg and P. J. Dagdigian}}()}]{hibridon}%
  \BibitemOpen
  \bibfield  {author} {\bibinfo {author} {\bibnamefont {{M. H. Alexander and D.
  E. Manolopoulos and H.-J. Werner and B. Follmeg and P. J. Dagdigian}}},\
  }\href {http://www2.chem.umd.edu/groups/alexander/hibridon/hib43/} {\enquote
  {\bibinfo {title} {Hibridon package},}\ }\BibitemShut {NoStop}%
\bibitem [{\citenamefont {Ziemkiewicz}\ \emph {et~al.}(2012)\citenamefont
  {Ziemkiewicz}, \citenamefont {Pluetzer}, \citenamefont {Nesbitt},
  \citenamefont {Scribano}, \citenamefont {Faure},\ and\ \citenamefont {van~der
  Avoird}}]{Ziemkiewicz:2012aa}%
  \BibitemOpen
  \bibfield  {author} {\bibinfo {author} {\bibfnamefont {M.~P.}\ \bibnamefont
  {Ziemkiewicz}}, \bibinfo {author} {\bibfnamefont {C.}~\bibnamefont
  {Pluetzer}}, \bibinfo {author} {\bibfnamefont {D.~J.}\ \bibnamefont
  {Nesbitt}}, \bibinfo {author} {\bibfnamefont {Y.}~\bibnamefont {Scribano}},
  \bibinfo {author} {\bibfnamefont {A.}~\bibnamefont {Faure}}, \ and\ \bibinfo
  {author} {\bibfnamefont {A.}~\bibnamefont {van~der Avoird}},\ }\href
  {\doibase {10.1063/1.4732581}} {\bibfield  {journal} {\bibinfo  {journal}
  {{JOURNAL OF CHEMICAL PHYSICS}}\ }\textbf {\bibinfo {volume} {{137}}}
  (\bibinfo {year} {{2012}}),\ {10.1063/1.4732581}}\BibitemShut {NoStop}%
\bibitem [{\citenamefont {{Valiron}}\ \emph {et~al.}(2008)\citenamefont
  {{Valiron}}, \citenamefont {{Wernli}}, \citenamefont {{Faure}}, \citenamefont
  {{Wiesenfeld}}, \citenamefont {{Rist}}, \citenamefont {{Ked{\v z}uch}},\ and\
  \citenamefont {{Noga}}}]{Valiron:2008aa}%
  \BibitemOpen
  \bibfield  {author} {\bibinfo {author} {\bibfnamefont {P.}~\bibnamefont
  {{Valiron}}}, \bibinfo {author} {\bibfnamefont {M.}~\bibnamefont {{Wernli}}},
  \bibinfo {author} {\bibfnamefont {A.}~\bibnamefont {{Faure}}}, \bibinfo
  {author} {\bibfnamefont {L.}~\bibnamefont {{Wiesenfeld}}}, \bibinfo {author}
  {\bibfnamefont {C.}~\bibnamefont {{Rist}}}, \bibinfo {author} {\bibfnamefont
  {S.}~\bibnamefont {{Ked{\v z}uch}}}, \ and\ \bibinfo {author} {\bibfnamefont
  {J.}~\bibnamefont {{Noga}}},\ }\href {\doibase 10.1063/1.2988314} {\bibfield
  {journal} {\bibinfo  {journal} {\jcp}\ }\textbf {\bibinfo {volume} {129}},\
  \bibinfo {pages} {134306} (\bibinfo {year} {2008})}\BibitemShut {NoStop}%
\bibitem [{\citenamefont {{Faure}}\ \emph {et~al.}(2005)\citenamefont
  {{Faure}}, \citenamefont {{Wiesenfeld}}, \citenamefont {{Wernli}},\ and\
  \citenamefont {{Valiron}}}]{Faure:2005aa}%
  \BibitemOpen
  \bibfield  {author} {\bibinfo {author} {\bibfnamefont {A.}~\bibnamefont
  {{Faure}}}, \bibinfo {author} {\bibfnamefont {L.}~\bibnamefont
  {{Wiesenfeld}}}, \bibinfo {author} {\bibfnamefont {M.}~\bibnamefont
  {{Wernli}}}, \ and\ \bibinfo {author} {\bibfnamefont {P.}~\bibnamefont
  {{Valiron}}},\ }\href {\doibase 10.1063/1.2033767} {\bibfield  {journal}
  {\bibinfo  {journal} {\jcp}\ }\textbf {\bibinfo {volume} {123}},\ \bibinfo
  {pages} {104309} (\bibinfo {year} {2005})}\BibitemShut {NoStop}%
\bibitem [{\citenamefont {Kyr{\"o}}(1981)}]{Kyro:1981aa}%
  \BibitemOpen
  \bibfield  {author} {\bibinfo {author} {\bibfnamefont {E.}~\bibnamefont
  {Kyr{\"o}}},\ }\href {\doibase https://doi.org/10.1016/0022-2852(81)90361-1}
  {\bibfield  {journal} {\bibinfo  {journal} {Journal of Molecular
  Spectroscopy}\ }\textbf {\bibinfo {volume} {88}},\ \bibinfo {pages} {167}
  (\bibinfo {year} {1981})}\BibitemShut {NoStop}%
\bibitem [{\citenamefont {{Manolopoulos}}(1986)}]{manolopoulos86}%
  \BibitemOpen
  \bibfield  {author} {\bibinfo {author} {\bibfnamefont {D.~E.}\ \bibnamefont
  {{Manolopoulos}}},\ }\href {\doibase 10.1063/1.451472} {\bibfield  {journal}
  {\bibinfo  {journal} {\jcp}\ }\textbf {\bibinfo {volume} {85}},\ \bibinfo
  {pages} {6425} (\bibinfo {year} {1986})}\BibitemShut {NoStop}%
\bibitem [{\citenamefont {{Baudry}}\ \emph
  {et~al.}(2018{\natexlab{b}})\citenamefont {{Baudry}}, \citenamefont
  {{Humphreys}}, \citenamefont {{Herpin}}, \citenamefont {{Torstensson}},
  \citenamefont {{Vlemmings}}, \citenamefont {{Richards}}, \citenamefont
  {{Gray}}, \citenamefont {{De Breuck}},\ and\ \citenamefont
  {{Olberg}}}]{Baudry:2018aa}%
  \BibitemOpen
  \bibfield  {author} {\bibinfo {author} {\bibfnamefont {A.}~\bibnamefont
  {{Baudry}}}, \bibinfo {author} {\bibfnamefont {E.~M.~L.}\ \bibnamefont
  {{Humphreys}}}, \bibinfo {author} {\bibfnamefont {F.}~\bibnamefont
  {{Herpin}}}, \bibinfo {author} {\bibfnamefont {K.}~\bibnamefont
  {{Torstensson}}}, \bibinfo {author} {\bibfnamefont {W.~H.~T.}\ \bibnamefont
  {{Vlemmings}}}, \bibinfo {author} {\bibfnamefont {A.~M.~S.}\ \bibnamefont
  {{Richards}}}, \bibinfo {author} {\bibfnamefont {M.~D.}\ \bibnamefont
  {{Gray}}}, \bibinfo {author} {\bibfnamefont {C.}~\bibnamefont {{De Breuck}}},
  \ and\ \bibinfo {author} {\bibfnamefont {M.}~\bibnamefont {{Olberg}}},\
  }\href {\doibase 10.1051/0004-6361/201731694} {\bibfield  {journal} {\bibinfo
   {journal} {\aap}\ }\textbf {\bibinfo {volume} {609}},\ \bibinfo {eid} {A25}
  (\bibinfo {year} {2018}{\natexlab{b}})},\ \Eprint
  {http://arxiv.org/abs/1711.02350} {arXiv:1711.02350 [astro-ph.SR]}
  \BibitemShut {NoStop}%
\bibitem [{\citenamefont {Yang}\ \emph {et~al.}(2006)\citenamefont {Yang},
  \citenamefont {Perera}, \citenamefont {Balakrishnan}, \citenamefont
  {Forrey},\ and\ \citenamefont {Stancil}}]{Yang:2006aa}%
  \BibitemOpen
  \bibfield  {author} {\bibinfo {author} {\bibfnamefont {B.}~\bibnamefont
  {Yang}}, \bibinfo {author} {\bibfnamefont {H.}~\bibnamefont {Perera}},
  \bibinfo {author} {\bibfnamefont {N.}~\bibnamefont {Balakrishnan}}, \bibinfo
  {author} {\bibfnamefont {R.~C.}\ \bibnamefont {Forrey}}, \ and\ \bibinfo
  {author} {\bibfnamefont {P.~C.}\ \bibnamefont {Stancil}},\ }\href {\doibase
  10.1088/0953-4075/39/19/s29} {\bibfield  {journal} {\bibinfo  {journal}
  {Journal of Physics B: Atomic, Molecular and Optical Physics}\ }\textbf
  {\bibinfo {volume} {39}},\ \bibinfo {pages} {S1229} (\bibinfo {year}
  {2006})}\BibitemShut {NoStop}%
\bibitem [{\citenamefont {{Daniel}}\ \emph {et~al.}(2010)\citenamefont
  {{Daniel}}, \citenamefont {{Dubernet}}, \citenamefont {{Pacaud}},\ and\
  \citenamefont {{Grosjean}}}]{Daniel:2010aa}%
  \BibitemOpen
  \bibfield  {author} {\bibinfo {author} {\bibfnamefont {F.}~\bibnamefont
  {{Daniel}}}, \bibinfo {author} {\bibfnamefont {M.~L.}\ \bibnamefont
  {{Dubernet}}}, \bibinfo {author} {\bibfnamefont {F.}~\bibnamefont
  {{Pacaud}}}, \ and\ \bibinfo {author} {\bibfnamefont {A.}~\bibnamefont
  {{Grosjean}}},\ }\href {\doibase 10.1051/0004-6361/200913745} {\bibfield
  {journal} {\bibinfo  {journal} {\aap}\ }\textbf {\bibinfo {volume} {517}},\
  \bibinfo {eid} {A13} (\bibinfo {year} {2010})}\BibitemShut {NoStop}%
\bibitem [{\citenamefont {{Bouhafs}}\ \emph {et~al.}(2019)\citenamefont
  {{Bouhafs}}, \citenamefont {{Rist}}, \citenamefont {{Daniel}}, \citenamefont
  {{Dumouchel}}, \citenamefont {{Lique}}, \citenamefont {{Wiesenfeld}},\ and\
  \citenamefont {{Faure}}}]{Bouhafs:2019aa}%
  \BibitemOpen
  \bibfield  {author} {\bibinfo {author} {\bibfnamefont {N.}~\bibnamefont
  {{Bouhafs}}}, \bibinfo {author} {\bibfnamefont {C.}~\bibnamefont {{Rist}}},
  \bibinfo {author} {\bibfnamefont {F.}~\bibnamefont {{Daniel}}}, \bibinfo
  {author} {\bibfnamefont {F.}~\bibnamefont {{Dumouchel}}}, \bibinfo {author}
  {\bibfnamefont {F.}~\bibnamefont {{Lique}}}, \bibinfo {author} {\bibfnamefont
  {L.}~\bibnamefont {{Wiesenfeld}}}, \ and\ \bibinfo {author} {\bibfnamefont
  {A.}~\bibnamefont {{Faure}}},\ }\href {\doibase 10.1093/mnras/sty3487}
  {\bibfield  {journal} {\bibinfo  {journal} {\mnras}\ }\textbf {\bibinfo
  {volume} {483}},\ \bibinfo {pages} {4639} (\bibinfo {year}
  {2019})}\BibitemShut {NoStop}%
\bibitem [{\citenamefont {Yang}\ \emph {et~al.}(2020)\citenamefont {Yang},
  \citenamefont {Zhang}, \citenamefont {Qu}, \citenamefont {Stancil},
  \citenamefont {Bowman}, \citenamefont {Balakrishnan},\ and\ \citenamefont
  {Forrey}}]{Yang:2020aa}%
  \BibitemOpen
  \bibfield  {author} {\bibinfo {author} {\bibfnamefont {B.}~\bibnamefont
  {Yang}}, \bibinfo {author} {\bibfnamefont {P.}~\bibnamefont {Zhang}},
  \bibinfo {author} {\bibfnamefont {C.}~\bibnamefont {Qu}}, \bibinfo {author}
  {\bibfnamefont {P.}~\bibnamefont {Stancil}}, \bibinfo {author} {\bibfnamefont
  {J.}~\bibnamefont {Bowman}}, \bibinfo {author} {\bibfnamefont
  {N.}~\bibnamefont {Balakrishnan}}, \ and\ \bibinfo {author} {\bibfnamefont
  {R.}~\bibnamefont {Forrey}},\ }\href {\doibase
  https://doi.org/10.1016/j.chemphys.2020.110695} {\bibfield  {journal}
  {\bibinfo  {journal} {Chemical Physics}\ }\textbf {\bibinfo {volume} {532}},\
  \bibinfo {pages} {110695} (\bibinfo {year} {2020})}\BibitemShut {NoStop}%
\bibitem [{\citenamefont {{Daniel}}, \citenamefont {{Dubernet}},\ and\
  \citenamefont {{Grosjean}}(2011)}]{Daniel:2011ab}%
  \BibitemOpen
  \bibfield  {author} {\bibinfo {author} {\bibfnamefont {F.}~\bibnamefont
  {{Daniel}}}, \bibinfo {author} {\bibfnamefont {M.~L.}\ \bibnamefont
  {{Dubernet}}}, \ and\ \bibinfo {author} {\bibfnamefont {A.}~\bibnamefont
  {{Grosjean}}},\ }\href {\doibase 10.1051/0004-6361/201118049} {\bibfield
  {journal} {\bibinfo  {journal} {\aap}\ }\textbf {\bibinfo {volume} {536}},\
  \bibinfo {eid} {A76} (\bibinfo {year} {2011})}\BibitemShut {NoStop}%
\bibitem [{\citenamefont {{Daniel}}\ \emph {et~al.}(2011)\citenamefont
  {{Daniel}}, \citenamefont {{Dubernet}}, \citenamefont {{Doronin}},
  \citenamefont {{Nenadovic}},\ and\ \citenamefont {{Bureau}}}]{Daniel:2011aa}%
  \BibitemOpen
  \bibfield  {author} {\bibinfo {author} {\bibfnamefont {F.}~\bibnamefont
  {{Daniel}}}, \bibinfo {author} {\bibfnamefont {M.~L.}\ \bibnamefont
  {{Dubernet}}}, \bibinfo {author} {\bibfnamefont {M.}~\bibnamefont
  {{Doronin}}}, \bibinfo {author} {\bibfnamefont {L.}~\bibnamefont
  {{Nenadovic}}}, \ and\ \bibinfo {author} {\bibfnamefont {J.}~\bibnamefont
  {{Bureau}}},\ }in\ \href@noop {} {\emph {\bibinfo {booktitle} {The Molecular
  Universe}}},\ Vol.\ \bibinfo {volume} {280},\ \bibinfo {editor} {edited by\
  \bibinfo {editor} {\bibfnamefont {J.}~\bibnamefont {{Cernicharo}}}\ and\
  \bibinfo {editor} {\bibfnamefont {R.}~\bibnamefont {{Bachiller}}}}\ (\bibinfo
  {year} {2011})\ p.\ \bibinfo {pages} {395}\BibitemShut {NoStop}%
\bibitem [{\citenamefont {{Faure}}\ and\ \citenamefont
  {{Josselin}}(2008)}]{Faure:2008aa}%
  \BibitemOpen
  \bibfield  {author} {\bibinfo {author} {\bibfnamefont {A.}~\bibnamefont
  {{Faure}}}\ and\ \bibinfo {author} {\bibfnamefont {E.}~\bibnamefont
  {{Josselin}}},\ }\href {\doibase 10.1051/0004-6361:200810717} {\bibfield
  {journal} {\bibinfo  {journal} {\aap}\ }\textbf {\bibinfo {volume} {492}},\
  \bibinfo {pages} {257} (\bibinfo {year} {2008})}\BibitemShut {NoStop}%
\bibitem [{\citenamefont {{Zittel}}\ and\ \citenamefont
  {{Masturzo}}(1991)}]{Zittel:1991aa}%
  \BibitemOpen
  \bibfield  {author} {\bibinfo {author} {\bibfnamefont {P.~F.}\ \bibnamefont
  {{Zittel}}}\ and\ \bibinfo {author} {\bibfnamefont {D.~E.}\ \bibnamefont
  {{Masturzo}}},\ }\href {\doibase 10.1063/1.461331} {\bibfield  {journal}
  {\bibinfo  {journal} {\jcp}\ }\textbf {\bibinfo {volume} {95}},\ \bibinfo
  {pages} {8005} (\bibinfo {year} {1991})}\BibitemShut {NoStop}%
\bibitem [{\citenamefont {{Barber}}\ \emph {et~al.}(2006)\citenamefont
  {{Barber}}, \citenamefont {{Tennyson}}, \citenamefont {{Harris}},\ and\
  \citenamefont {{Tolchenov}}}]{Barber:2006aa}%
  \BibitemOpen
  \bibfield  {author} {\bibinfo {author} {\bibfnamefont {R.~J.}\ \bibnamefont
  {{Barber}}}, \bibinfo {author} {\bibfnamefont {J.}~\bibnamefont
  {{Tennyson}}}, \bibinfo {author} {\bibfnamefont {G.~J.}\ \bibnamefont
  {{Harris}}}, \ and\ \bibinfo {author} {\bibfnamefont {R.~N.}\ \bibnamefont
  {{Tolchenov}}},\ }\href {\doibase 10.1111/j.1365-2966.2006.10184.x}
  {\bibfield  {journal} {\bibinfo  {journal} {\mnras}\ }\textbf {\bibinfo
  {volume} {368}},\ \bibinfo {pages} {1087} (\bibinfo {year} {2006})},\ \Eprint
  {http://arxiv.org/abs/astro-ph/0601236} {arXiv:astro-ph/0601236 [astro-ph]}
  \BibitemShut {NoStop}%
\bibitem [{\citenamefont {Wang}\ \emph {et~al.}(2020)\citenamefont {Wang},
  \citenamefont {Zhang}, \citenamefont {Zhai}, \citenamefont {Nooijen},\ and\
  \citenamefont {Li}}]{Wang:2020aa}%
  \BibitemOpen
  \bibfield  {author} {\bibinfo {author} {\bibfnamefont {L.}~\bibnamefont
  {Wang}}, \bibinfo {author} {\bibfnamefont {X.-L.}\ \bibnamefont {Zhang}},
  \bibinfo {author} {\bibfnamefont {Y.}~\bibnamefont {Zhai}}, \bibinfo {author}
  {\bibfnamefont {M.}~\bibnamefont {Nooijen}}, \ and\ \bibinfo {author}
  {\bibfnamefont {H.}~\bibnamefont {Li}},\ }\href {\doibase 10.1063/5.0009098}
  {\bibfield  {journal} {\bibinfo  {journal} {The Journal of Chemical Physics}\
  }\textbf {\bibinfo {volume} {153}},\ \bibinfo {pages} {054303} (\bibinfo
  {year} {2020})},\ \Eprint
  {http://arxiv.org/abs/https://doi.org/10.1063/5.0009098}
  {https://doi.org/10.1063/5.0009098} \BibitemShut {NoStop}%
\bibitem [{\citenamefont {Santos}\ \emph {et~al.}(2011)\citenamefont {Santos},
  \citenamefont {Balakrishnan}, \citenamefont {Lepp}, \citenamefont
  {Qu{\'e}m{\'e}ner}, \citenamefont {Forrey}, \citenamefont {Hinde},\ and\
  \citenamefont {Stancil}}]{Santos:2011aa}%
  \BibitemOpen
  \bibfield  {author} {\bibinfo {author} {\bibfnamefont {S.~F.~d.}\
  \bibnamefont {Santos}}, \bibinfo {author} {\bibfnamefont {N.}~\bibnamefont
  {Balakrishnan}}, \bibinfo {author} {\bibfnamefont {S.}~\bibnamefont {Lepp}},
  \bibinfo {author} {\bibfnamefont {G.}~\bibnamefont {Qu{\'e}m{\'e}ner}},
  \bibinfo {author} {\bibfnamefont {R.~C.}\ \bibnamefont {Forrey}}, \bibinfo
  {author} {\bibfnamefont {R.~J.}\ \bibnamefont {Hinde}}, \ and\ \bibinfo
  {author} {\bibfnamefont {P.~C.}\ \bibnamefont {Stancil}},\ }\href {\doibase
  10.1063/1.3595134} {\bibfield  {journal} {\bibinfo  {journal} {The Journal of
  Chemical Physics}\ }\textbf {\bibinfo {volume} {134}},\ \bibinfo {pages}
  {214303} (\bibinfo {year} {2011})},\ \Eprint
  {http://arxiv.org/abs/https://doi.org/10.1063/1.3595134}
  {https://doi.org/10.1063/1.3595134} \BibitemShut {NoStop}%
\end{thebibliography}%

\end{document}